\documentclass[copyright]{eptcs}
\providecommand{\event}{Component and Service Interoperability} % Name of the event you are submitting to

\usepackage{amssymb}

\usepackage{subfigure}

\usepackage{graphicx}

\newcommand {\ie} {{\em i.e.}}
\newcommand {\eg} {{\em e.g.}}
\newcommand {\etal} {{\em et al.\ }}

\newcommand{\trans}[1]{\stackrel{#1}{\longrightarrow}}

\newcommand{\wtrans}[1]{\stackrel{#1}{\Longrightarrow}}

\usepackage[normalem]{ulem} % for \sout
\usepackage{xcolor}

\usepackage{ifthen}
\usepackage{amssymb}
\newboolean{showcomments}
\setboolean{showcomments}{true} % toggle to show or hide comments
\ifthenelse{\boolean{showcomments}}
  {\newcommand{\nb}[2]{
    \fcolorbox{gray}{yellow}{\bfseries\sffamily\scriptsize#1}
    {\sf\small$\blacktriangleright$\textit{#2}$\blacktriangleleft$}
   }
   
  }
  {\newcommand{\nb}[2]{}
   
  }

\title{Partition Refinement of Component Interaction Automata: Why Structure Matters More Than Size}

\author{Markus Lumpe and Rajesh Vasa
\institute{Faculty of Information \& Communication Technologies\\
Swinburne University of Technology\\
Hawthorn, Australia}
\email{\{mlumpe,rvasa\}@swin.edu.au}
}

\def\titlerunning{Partition Refinement of Component Interaction Automata: Why Structure Matters More Than Size}

\def\authorrunning{M. Lumpe and R. Vasa}

\newtheorem{definition}{Definition}

\begin{document}

\maketitle

\begin{abstract}
Automata-based modeling languages, like \emph{Component Interaction Automata}, offer an attractive means to capture and analyze the behavioral aspects of interacting components. At the center of these modeling languages we find \emph{finite state machines} that allow for a fine-grained description how and when specific service requests may interact with other components or the environment. Unfortunately, automata-based approaches suffer from exponential state explosion, a major obstacle to the successful application of these formalisms in modeling real-world scenarios. In order to cope with the complexity of individual specifications we can apply \emph{partition refinement}, an abstraction technique to alleviate the state explosion problem. But this technique too exhibits exponential time and space complexity and, worse, does not offer any guarantees for success. To better understand as to why partition refinement  succeeds in some cases while it fails in others, we conducted an empirical study on the performance of a partition refinement algorithm for \emph{Component Interaction Automata} specifications. As a result we have identified suitable \emph{predictors} for the expected effectiveness of partition refinement. It is the structure, not the size, of a specification that weighs heavier on the outcome of partition refinement. In particular, \emph{Component Interaction Automata} specifications for real-world systems are capable of producing \emph{scale-free networks} containing structural artifacts that can assist the partition refinement algorithm not only converge earlier,  but also yield a significant state space reduction on occasion.
\end{abstract}

% start body

% Intro.tex

\section{Introduction}

\emph{Component Interaction Automata} \cite{brim:06a,cerna:07a} offers a well-balanced formal modeling framework to capture both the temporal and the hierarchical aspects of cooperating components in modern real-world component-oriented software systems. The \emph{Component Interaction Automata} formalism provides two component-oriented software development processes: the \emph{architectural description} of the system being developed and the \emph{formal verification} of the intrinsic properties of the system under consideration \cite{brim:06a}. The \emph{Component Interaction Automata} modeling language builds on \emph{I/O Automata} \cite{lynch:87a}, \emph{Interface Automata} \cite{alfaro:01a}, and \emph{Team Automata} \cite{beek:03a} that all employ an \emph{automata-based language} to represent the assumptions about a system's capabilities to interact with the environment or other components. However, unlike its predecessors, \emph{Component Interaction Automata} distinguishes between components and component instances \cite{lumpe:99a}. This embodies a crucial difference that makes the \emph{Component Interaction Automata} approach more suitable for the specification of real-world component-oriented systems \cite{lumpe:10b}.

Unfortunately, automata-based modeling approaches suffer from \emph{combinatorial state space explosion} with respect to the size of the modeled system. When defining the composition of components, we need to construct the \emph{product automaton} \cite{hopcroft:07a} of the system being specified. Even though not all states in the product automaton may be reachable (\ie, they can be removed from the system), composite component interaction automata will eventually grow to a size where an effective analysis of the system properties may not be feasible \cite{lumpe:08a}.

It is for this reason that we have been studying suitable abstraction mechanisms in order to distill smaller, yet behaviorally equivalent, specifications for a given component interaction automaton. In particular, we have developed a bisimulation-based \emph{partition refinement} algorithm for \emph{Component Interaction Automata} \cite{lumpe:08a,lumpe:10b}. Partition refinement \cite{hermanns:02a} constructs, if possible, a new image of a given automaton, where the states of the new automaton correspond to the \emph{equivalence classes} of the old automaton. The granularity of the refinement process depends on the underlying equivalence relation being used. For \emph{Component Interaction Automata} we use \emph{weak bisimluation}, a behavioral equivalence relation that abstracts from internal component synchronizations. In other words, partition refinement for \emph{Component Interaction Automata} equates both behavioral equivalent substructures of an automaton and states that are solely connected by internal component synchronizations \cite{lumpe:08a,lumpe:10b}.

\begin{figure}[h]
\centering
\subfigure[][Composite automaton]{
\includegraphics[width=0.38\textwidth]{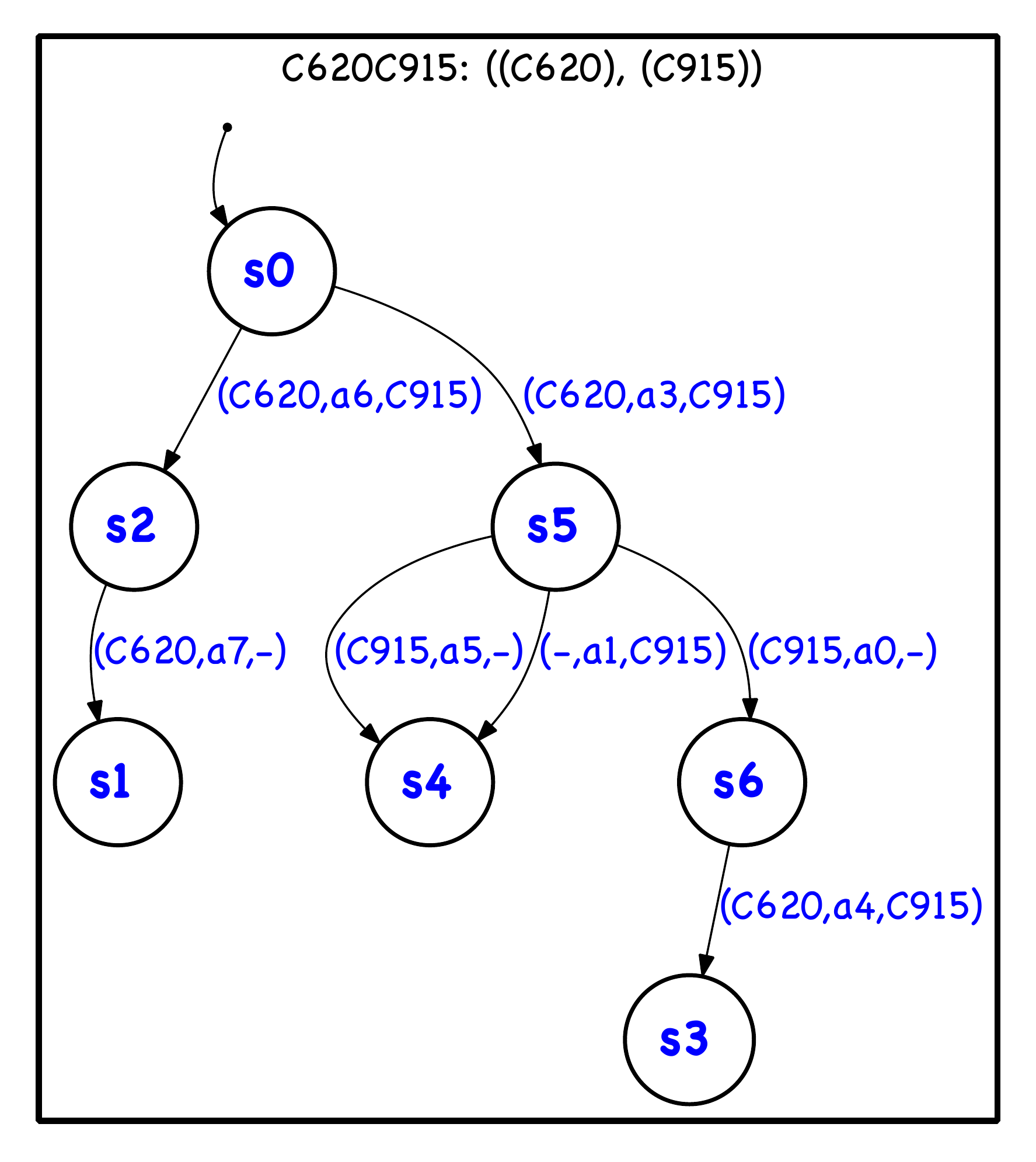}
\label{fig1A.fig}
}
\hspace{2mm}
\subfigure[][Reduced automaton]{
\includegraphics[width=.38\textwidth]{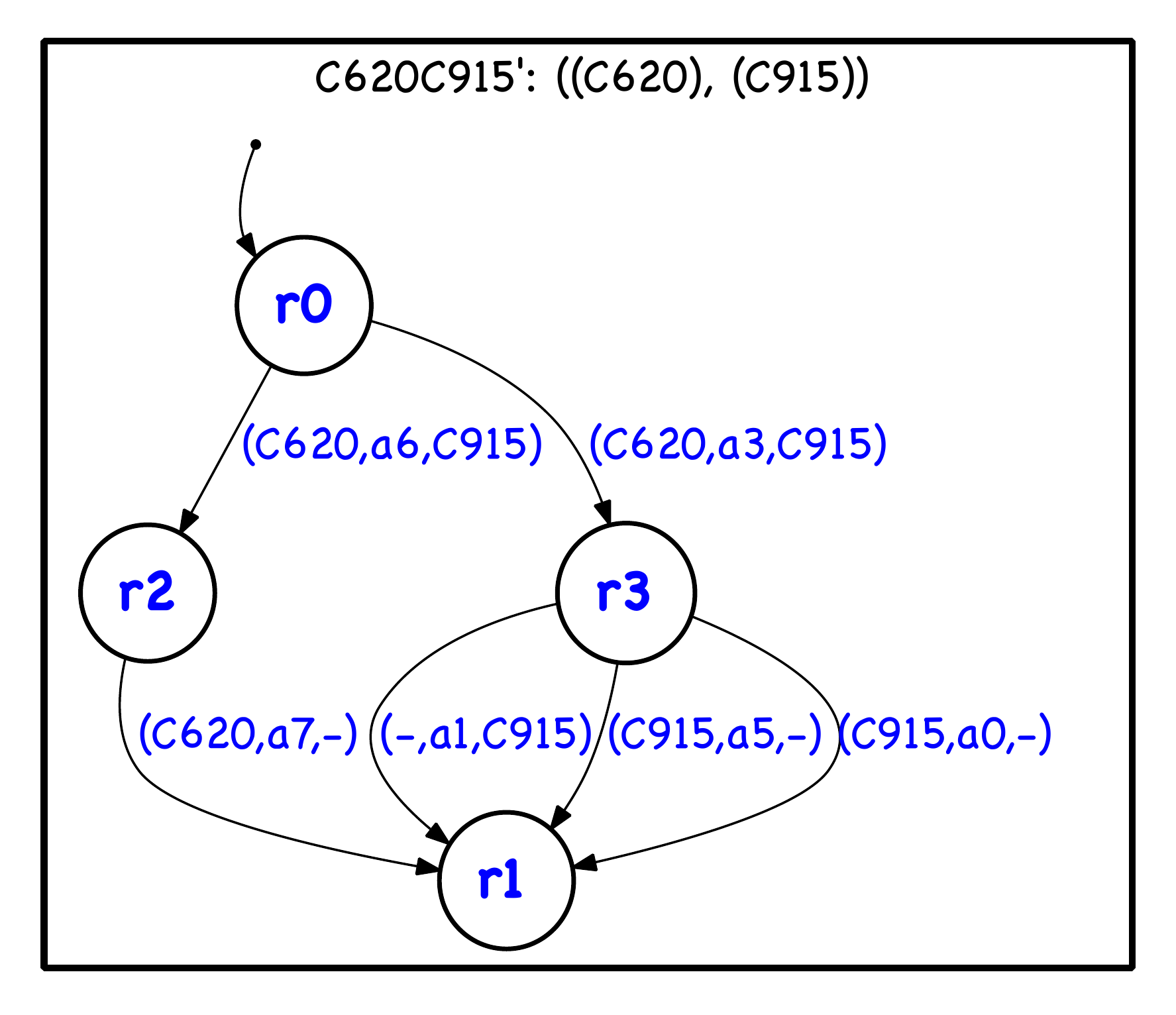}
\label{fig1B.fig}
}
\label{fig1.fig}
\caption{Component interaction automaton $C620C915$ and its reduced variant $C620C915'$.}
\end{figure}

Consider, for example, the graphical representation of the automata $C620C915$ and $C620915'$ shown in Figure~\ref{fig1.fig}. Automaton $C620C915$ is drawn from a sample of experimental components interaction automata specifications used in our study. The states $s1$, $s3$, $s4$, and $s6$ in $C620C915$ are \emph{weakly-bisimilar} and belong, therefore, to the same equivalence class denoted by state $r1$ in automaton $C620C915'$ (cf. Figure~\ref{fig1B.fig}). This small example illustrates two specific properties of \emph{Component Interaction Automata} and partition refinement. First, partition refinement through weak bisimulation does not remove all internal component synchronizations. Both $(C620,a6,C915)$ and $(C620,a3,C915)$ have to remain in $C620C915'$ as their respective target states offer different interaction capabilities. Second, the states $s1$, $s3$, $s4$, and $s6$ in $C620C915$ form a \emph{community} \cite{fortunato:09a} or \emph{synchronization clique} \cite{lumpe:10b} that gives rise to a significant reduction. Here, the reduction involves a terminal state, but if $C620C915'$ were to occur within a larger system, this  particular effect would enable the partition refinement algorithm to yield a better reduction ratio. 

The presence of synchronization cliques in an automaton is of particular significance for the understanding of the performance of the partition refinement algorithm, as the refinement process itself is not guaranteed to succeed. It order to identify the reasons as to why partition refinement can sometimes yield strong state space reduction ratios \cite{lumpe:08a}, while it fails completely on other occasions, we have run an analysis on a sample of 1,680 experimental composite systems. Each system consists of between 2 and 11 machine-generated \emph{Component Interaction Automata} specifications that all enjoy topological properties similar to real-world software systems \cite{vasa:05a,vasa:07a}. Every experiment was allowed to run at most two hours and was carried out on a Mac Pro equipped with one 2.66 GHz Quad-Core processor and 8GB 1066 MHz DDR3 memory running Mac OS X 10.6.3. The results were analyzed using \emph{logistic regression} \cite{agresti:02a,boslaugh:08a}, a statistical method for the prediction of the probability of the occurrence of a specific event. In particular, we wanted to determine which features of a \emph{Component Interaction Automata} specification can serve as \emph{explanatory variables} or \emph{predictors} for a specific expected outcome of running partition refinement on a given component interaction automaton. By identifying suitable explanatory variables we can construct a \emph{model} that explains how and when partition refinement for a \emph{Component Interaction Automata} specification is to succeed or fail.

\smallskip

The rest of the paper is organized as follows: in Section~\ref{formalism.sec} we briefly review the \emph{Component Interaction Automata} formalism and present a corresponding partition refinement algorithm.  We proceed with an analysis of the structural properties of \emph{Component Interaction Automata} specifications in Section~\ref{structure.sec}. In particular, we study selected graph properties  and highlight how they affect the possible outcome of partition refinement. Section~\ref{regression.sec} presents the results of our logistic regression analysis. In particular, we discuss four models using \emph{maximum likelihood estimation} (MLE) and demonstrate that structure, not size, provides good estimates for the success of partition refinement. We conclude with a summary of main observations in Section~\ref{conc.sec}.

% Formalism.tex

\section{Partition Refinement for Component Interaction Automata}
\label{formalism.sec}

The \emph{Component Interaction Automata} formalism aims at the specification and verification of component-based software systems at an interface level \cite{brim:06a,cerna:07a}. It is interfaces that allow us to define a suitable decomposition of a system into its logical units, the \emph{components}, and that capture the components' interactive behavior in a concise way. Collectively, interfaces and the information they relay form a \emph{contractual specification} \cite{beugnard:99} that explicitly states all assumptions about a component's (or system's) deployment environment. Using the \emph{Component Interaction Automata} formalism we can reason about contractual specifications in at least two ways: \emph{``Does the system respond to service requests in the expected order?"} and \emph{``What is a behaviorally equivalent specification for a given component or system?"}

\begin{definition}[Component Interaction Automata]\label{cia.def}
A component interaction automaton $\mathcal{C}$ is a quintuple $(Q,
\mathit{Act}, \delta, I, H)$ where:
\begin{itemize}
\item[$\bullet$] 
$Q$ is a finite set of states,

\item[$\bullet$]
$\mathit{Act}$ is a finite set of actions,

\item[$\bullet$]
$\delta \subseteq Q \times \Sigma \times Q$ is a finite set of labeled transitions, where
$\Sigma \subseteq \{ (S(H) \cup \{-\} \times \mathit{Act} \times S(H) \cup \{-\}) \} \setminus$ $\{(\{-\}\times \mathit{Act} \times \{-\})\}$ is the set of structured labels induced by $\mathcal{C}$,

\item[$\bullet$]
$I \subseteq Q$ is a non empty set of initial states, and

\item[$\bullet$]
$H$ is a hierarchical composition structure with either
\begin{itemize}
\item[$\bullet$]
$H = (C_1, ..., C_n)$ denoting a primitive composition of the component instances $C_1, ..., C_n$, such that $S(H) = \cup_{i=1}^n \{C_i\}$, or

\item[$\bullet$]
$H = (H_1, ..., H_m)$, where $H_1, ..., H_m$ are hierarchies of component instances satisfying the structural property $\forall\ 1 \leq i,j \leq m, i \not= j:\ S(H_i) \cap S(H_j) = \emptyset$, such that $S(H) = \cup_{i=1}^m S(H_i)$. 
\end{itemize}

\end{itemize}

\noindent
Each component interaction automaton is further characterized by two sets $P \subseteq \mathit{Act}$, the provided actions, and $R \subseteq \mathit{Act}$, the required actions. These sets capture the automaton's enabled interface with an environment. We write $\mathcal{C}^P_R$ to denote an automaton $\mathcal{C}$ that is input-enabled in $P$ and output-enabled in $R$.\hfill$\Box$
\end{definition}

The composition of component interaction automata is defined in the usual way. The behavior of the composite system is the cross-product of its component behaviors. We apply the architectural constraints $P$ and $R$, the set of provided services and the set of required services, respectively, to control, which transitions can occur in the composite automaton. In general, we use $P$ and $R$ to contain only those actions that appear in input and output transitions of the composite automaton.

\begin{definition}[Component Interaction Automata Composition]\label{comp.def}
Let $\mathcal{S}_{R}^{P} = \{ (Q_i, \mathit{Act}_i, \delta_i, I_i, H_i) \}_{i \in \mathbb{Z}}$ be a system of pairwise disjoint component interaction automata and $P,R$ are the provided and required actions. Then $\mathcal{C}^{P}_{R} = (\prod_{i}Q_i, \cup_{i}\mathit{Act}_i, \delta_{\mathcal{S}^{P}_{R}}, \prod_{i}I_i, (H_i)_{i})$
is the composite component interaction automaton of $\mathcal{S}_{R}^{P}$ where
$q_j$ denotes a function $\prod_{i}Q_i \rightarrow Q_j$, the projection from product state $q$ to the $j^\mathit{th}$ component's state $q$, and
\begin{center}
$\delta_{\mathcal{S}^{P}_{R}} = \delta_{OldSync} \cup \delta_{\mathit{NewSync}} \cup \delta_{\mathit{Input}} \cup \delta_{\mathit{Output}}$
\end{center}
with
\begin{eqnarray*}
\delta_{OldSync} & =  &\{ (q, (n_1, a, n_2), q')\ |\ \exists i: (q_i, (n_1, a, n_2), q_{i}') \in \delta_i\ \wedge\  \forall j \in \mathcal{I}, j \not= i:\ q_j = q_{j}' \},
\\
\hspace{0.4cm}
\delta_{\mathit{NewSync}} & = & \{  (q, (n_1, a, n_2), q')\ |\ \exists i_1, i_2\ \wedge\ 
i_1 \not= i_2: (q_{i_1}, (n_1, a, -), q_{i_1}') \in \delta_{i_1} \wedge\\[-1mm]
& & 
\hspace{4cm} 
(q_{i_2}, (-, a, n_2), q_{i_2}') \in \delta_{i_2} \wedge\ \forall j\ \wedge\ 
i_1 \not= j \not= i_2: 
q_j = q_{j}' \},
\\
\delta_{\mathit{Input}} & = & \{ (q, (-, a, n), q')\ |\ a \in R\ \wedge\ \exists i: 
(q_i, (-, a, n), q_{i}') \in \delta_i\ \wedge\ 
  \forall j\ \wedge\ j \not= i: q_j = q_{j}' \},
\\
\delta_{\mathit{Output}} & = & \{  (q, (n, a, -), q')\ |\ a \in P\ \wedge\ \exists i: 
(q_i, (n, a, -), q_{i}') \in \delta_i\ \wedge\ 
\forall j\ \wedge\ j \not= i: q_j = q_{j}' \}.\hspace{0.9cm}\Box
\end{eqnarray*}
\end{definition}

Composition in \emph{Component Interaction Automata} is defined over an arbitrary number of components. The behaviors of the individual components are simultaneously recombined to yield the composite behavior. This flexibility comes, however, at a price. The effects of the exponential combinatorial time and space explosion appear rather quickly \cite{lumpe:08a} and the resources required to build a product automaton exceed practical limits. For this reason we compose only two automata at a time and apply partition refinement to the result immediately in our experiments. This approach remains faithful to the underlying specification, but it provides us with a scenario in which we can think of partition refinement as an ``on-the-fly'' technique. 

In order to apply partition refine to a \emph{Component Interaction Automata} specification we need to define a suitable  equivalence relation. We use bisimulation, in particular \emph{weak bisimulation}, for this purpose. Weak bisimulation provides us with an equivalence relation that equates automata that only differ in the lengths of occurring internal component synchronization sequences \cite{cerna:07a,lumpe:10b}.

\begin{definition}[Weak Bisimulation for Component Interaction Automata]
\label{bisim.def}
Given two component interaction automata $A = (Q_A, \mathit{Act}_A, \delta_A, I_A, H)$ and $B = (Q_B, \mathit{Act}_B, \delta_B, H)$ with an identical composition hierarchy $H$, a binary relation $\mathcal{R} \subseteq Q \times Q$ with $Q =  Q_A \cup Q_B$ is a weak bisimulation, if it is symmetric and $(q,p) \in \mathcal{R}$ implies, for all $l \in \Sigma$, $\Sigma = \Sigma_A \cup \Sigma_B$ being the set of structured labels induced by $A$ and $B$,
\begin{enumerate}
\item[$\bullet$]
whenever $q \trans{l} q'$, then $\exists p'$ such that $p \wtrans{l} p'$ and $(q',p') \in \mathcal{R}$.
\end{enumerate}
Two component interaction automata $A$ and $B$ are weakly bisimilar, written
$A \approx B$, if they are related by some weak bisimulation. \hfill$\Box$
\end{definition}

We write $p \wtrans{l} p'$ to denote that an automaton, $\mathcal{C} = (Q, \mathit{Act}, \delta, I, H)$, can evolve from state $p$ to $p'$ through an interaction $l$ with a possibly empty sequence of internal component synchronizations occurring before and after transition $l$. The relation $p \wtrans{l} p'$ gives rise to a {\em splitter function} that provides us with a means to compute the equivalence classes up to weak bisimulation for a given automaton $\mathcal{C}$. The splitter function is a Boolean predicate $\gamma : Q \times \Sigma \times \mathbb{S} \mapsto \{\mathtt{true}, \mathtt{false}\}$, where $\mathbb{S} \subseteq 2^Q$ is a set of candidate equivalence classes for $\mathcal{C}$. Let $q$ be a state, $P \in \mathbb{S}$ be candidate equivalence class,  and $l$ be a structured label for a component interaction automaton $\mathcal{C}$ to be refined. Then the corresponding splitter is  
\begin{eqnarray}
\gamma(q,l,P) & := &
\left\{
\begin{array}{ll}
\mathtt{true}  & \mathit{if\ there\ is\ p \in P\ such\ that\ q \wtrans{l} p},\\
\mathtt{false} & \mathit{otherwise}
\end{array}
\right.
\end{eqnarray}

Partition refinement is a function $\mathbb{S} \times \Sigma \times \mathbb{S} \mapsto \mathbb{S}$ that takes three arguments: $X_{i-1} \in \mathbb{S}$, the partition resulting from step $i-1$, $l \in \Sigma$, the splitter label, and $P_i \in \mathbb{S}$, the equivalence classes in step $i$.  
\begin{eqnarray}
\mathit{refine}(X_{i-1},l,P_i) & := & 
\cup_{X \in X_{i-1}} ( \cup_{v \in \{ \mathtt{true}, \mathtt{false} \}} 
\{ q\ |\ \forall q \in X.\ \gamma(q,l,P_i) = v \} )  - \{ \emptyset \}
\end{eqnarray}

\newcommand{\kw}[1] {\textbf{#1}}

Our partition refinement algorithm differs in two aspects compared with the one
proposed by Hermanns \cite{hermanns:02a}. First, we add an iteration over the labels. Experiments have shown that the partition refinement algorithm will require fewer splitters if we add this extra iteration. Furthermore, we split $X$ in two sets: $X^{1}$, the set of singleton partitions, and $X^{>1}$, the set of partitions comprising two or more states. Singleton partitions cannot be further refined and, therefore, we do not need to test them again. The partition refinement algorithm has to test only $X^{>1}$:
{
\begin{tabbing}
\hspace{4.0cm}\=\hspace{0.4cm}\=\hspace{0.4cm}\=\hspace{0.4cm}\=\hspace{0.4cm}\=\hspace{0.4cm}\=\hspace{0.4cm}\=\hspace{0.8cm}\=\hspace{0.8cm}\=\kill
\> $X^{>1}_{i-1} \leftarrow \{Q\}$; $X^1_{i-1} \leftarrow \emptyset$; Repeat $\leftarrow \mathit{true}$; \\
\> \kw{while} Repeat  \\
\> \> \kw{do} \kw{for} $l \in \Sigma$ \\
\> \> \> \kw{do} EqvClasses $\leftarrow X^{>1}_{i-1} \cup X^1_{i-1}$; Repeat $\leftarrow \mathit{false}$; \\
\> \> \> \> \kw{while} EqvClasses $\not= \emptyset$  \\
\> \> \> \> \> \kw{do} \kw{choose} $P_i \in$ EqvClasses; \\
\> \> \> \> \> \> $(X^{>1}_i,X^{1}_i) \leftarrow \mathit{refine}(X^{>1}_{i-1},l,P_i)$; \\
\> \> \> \> \> \> \kw{if} $X^{>1}_i \not= X^{>1}_{i-1}$ \\ 
\> \> \> \> \> \> \> \kw{then} \> $X^{>1}_{i-1} \leftarrow X^{>1}_i$;
                                 $X^1_{i-1} \leftarrow X^1_{i-1} \cup X^1_i$; \\
\> \> \> \> \> \> \> \> EqvClasses $\leftarrow X^{>1}_{i-1} \cup X^1_{i-1}$; \\
\> \> \> \> \> \> \> \> Repeat $\leftarrow \mathit{true}$; \\
\> \> \> \> \> \> \> \kw{else} EqvClasses $\leftarrow$ EqvClasses $- \{ P_i \}$; \\
\> \kw{return} $X^{>1}_{i-1} \cup X^1_{i-1}$; 
\end{tabbing}
}

The above algorithm yields a partition that is minimal up to weak bisimulation (\ie, the fixed-point) with respect to the number of required states for a given automaton $\mathcal{C}$. This algorithm is part of our experimental composition framework for {\em Component Interaction Automata}, implemented in PLT-Scheme, that provides support not only for the specification and refinement of component interaction automata, but also for the extraction of metrics data \cite{lumpe:08a,lumpe:10b}.

% Structure.tex

\section{Structural Analysis}
\label{structure.sec}

The state explosion problem in \emph{Component Interaction Automata} is intrinsic to all algebraic software modeling techniques that seek to express properties of the modeled system through an automata-based approach. The actual specifications yield \emph{directed graphs} in which vertices are the states of the systems and edges denote possible interactions with the environment or other components. Graph theory \cite{fortunato:09a,newman:03a} provides a rich source for a meaningful interpretation of properties of \emph{Component Interaction Automata} specifications. However, even though partition refinement explores the communication structure of an automaton, partition refinement itself does not actually exploit the topology of the automaton's graph to fine-tune the refinement process. 

\begin{figure}[h]
\centering
\subfigure[][Experimental automaton C620]{
\includegraphics[width=0.29\textwidth]{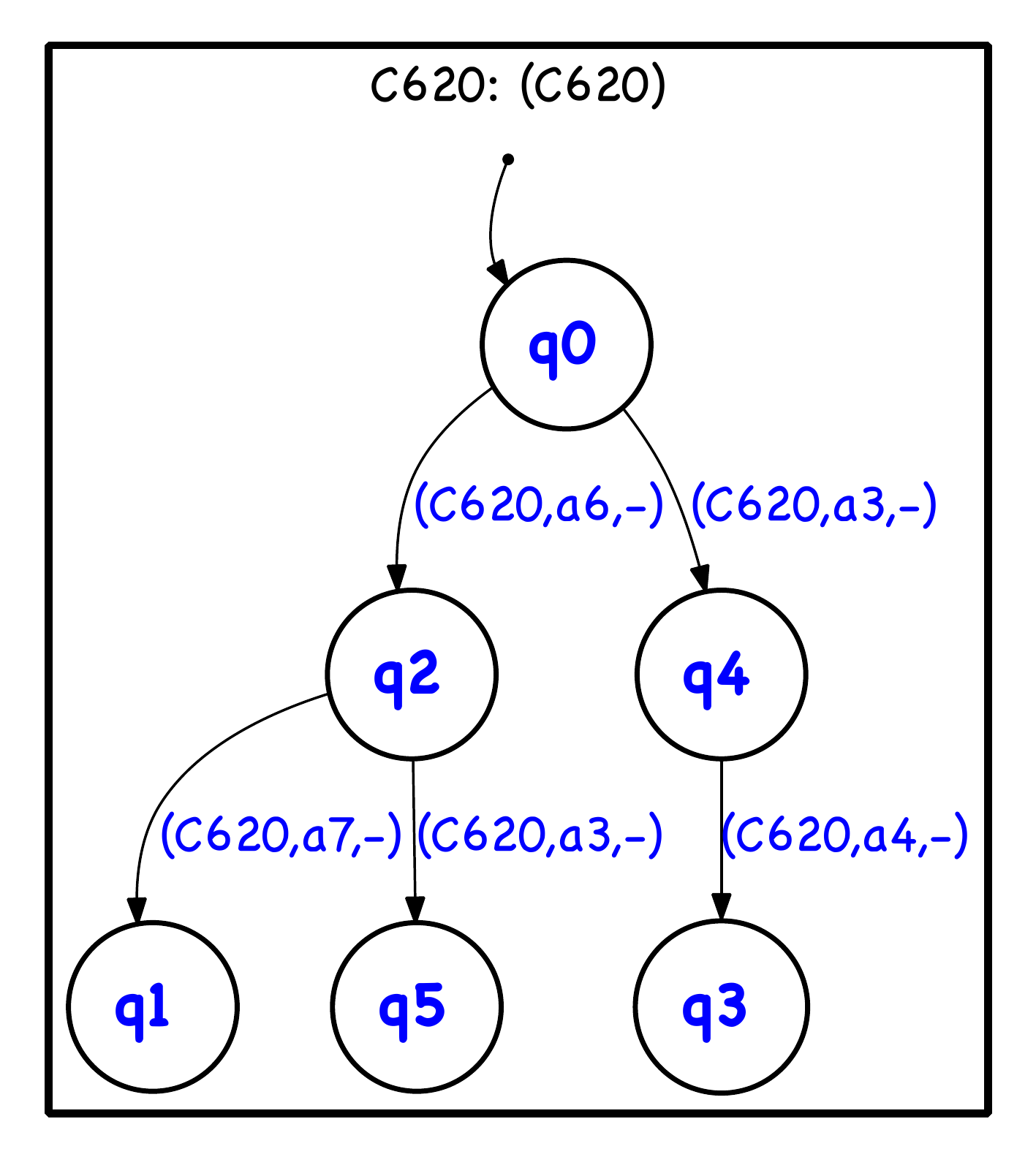}
\label{fig2A.fig}
}
\hspace{10mm}
\subfigure[][Experimental automaton C915]{
\includegraphics[width=0.29\textwidth]{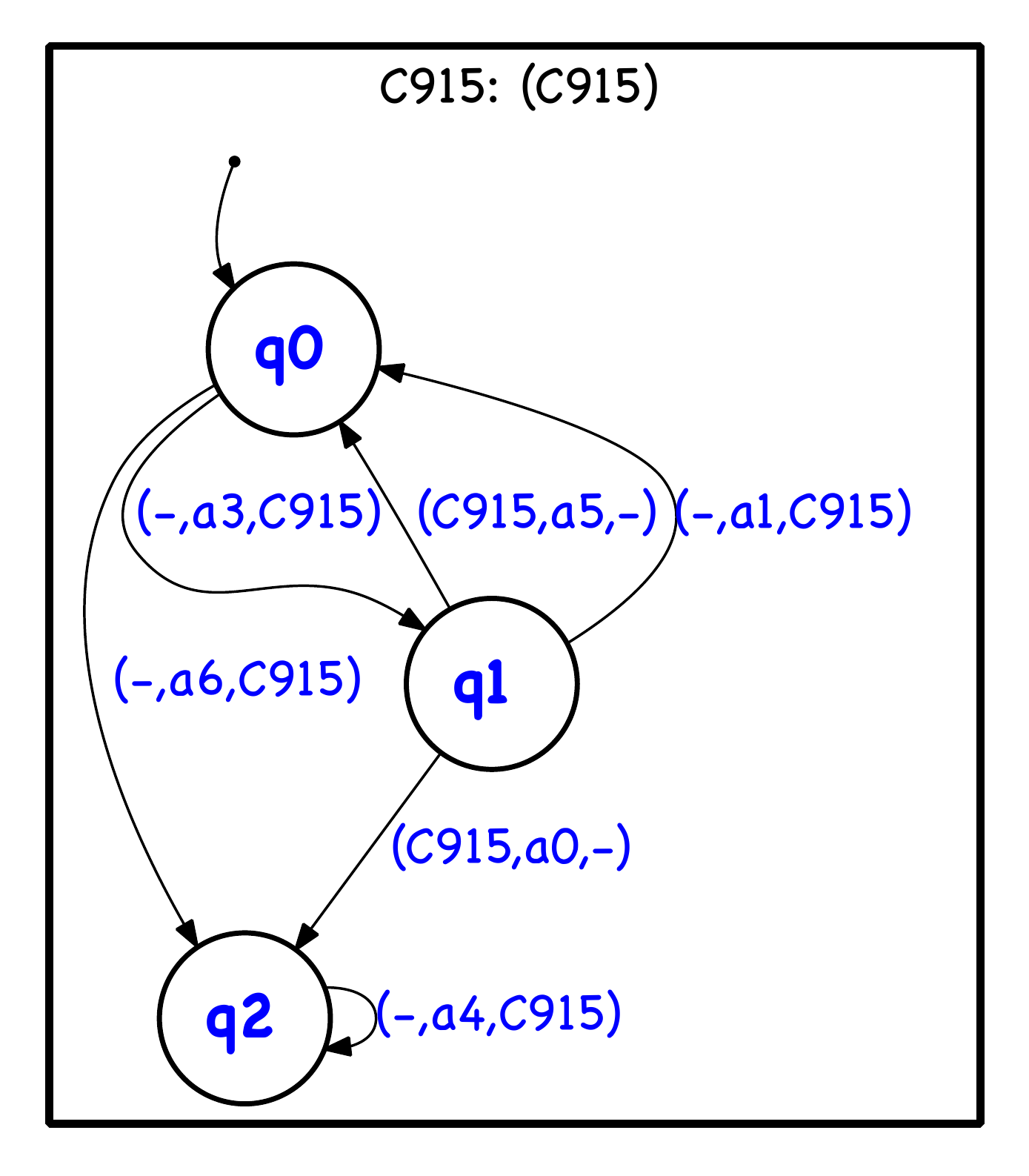}
\label{fig2B.fig}
}
\caption{Emerging preferential attachment in \emph{Component Interaction Automata}.}
\label{c620c915.fig}
\end{figure}

Consider the two automata, $C620$ and $C915$, shown in Figure~\ref{c620c915.fig}. These are two machine-generated specifications to simulate real-world software systems. Both automata exhibit some typical graph properties that we find in real-world software systems. First, software is not made of ``Lego blocks" \cite{potanin:05}. The topology of both automata varies greatly. The distribution of transitions in different automata does not follow a uniform pattern. Some states attract more transitions than others. There is no unique size in terms of number of states and number of transitions. Nevertheless, the ratio between both quantities assumes some common value, a feature that becomes even more pronounced when we compose automata specifications and refine the resulting composite.

Second, the automata $C620$ and $C915$ exhibit \emph{preferential attachment} \cite{barabasi:99a}. States that are already well connected attract new transitions more easily than others. This \emph{``rich-get-richer"} strategy is typical for software systems \cite{vasa:09a}. The distribution of functionality in a software system is neither regular nor random. Developers prefer to organize and maintain software systems around a small number of highly complex abstractions \cite{vasa:09a}. These abstractions constitute virtual \emph{hubs} in the systems and appear to guarantee not only the proper function of a software, but also the ability to evolve a software system in order to meet changing requirements in the future \cite{vasa:09a}. 

We can measure these structural features using two concepts: the \emph{scaling exponent} $\beta$ \cite{vasa:05a,vasa:07a} to denote the power-scaling relationship between the number of states and the number of transitions of an automaton and the \emph{Gini coefficients} \cite{gini:21a,vasa:09a} of the incoming and outgoing transitions in an automaton in order to quantify the degree of \emph{inequality in the distribution} of these attributes in a given automaton. Both measures provide suitable summary metrics  of the underlying directed graph topology of an automaton. Moreover, these measures can also serve as reliable predictors for the success of partition refinement. Hence, the better we understand the topological properties of a given automaton the more we can guide the partition refinement process, if possible, to yield a significant state space reduction.

The power-scaling relationship for a component interaction automaton $\mathcal{C} = (Q, \mathit{Act}, \delta, I, H)$ between the number of states and the number of transitions is given by
\begin{eqnarray}
|\delta| & \sim & |Q|^\beta
\end{eqnarray}
where the scaling exponent $\beta$ is the ratio between the natural logarithm of the number of states and the number of transitions in automaton $\mathcal{C}$: 
\begin{eqnarray}
\beta & = & \frac{\displaystyle{\mathrm{ln}(|\delta|)}}{\displaystyle{\mathrm{ln}(|Q|)}}
\end{eqnarray}

The actual value of the scaling exponent $\beta$ for our sample of 1,680 machine-generated automata specifications satisfies the probability density function 
\begin{eqnarray}
P[a \le X \le b] = \int_{a}^{b} f(x)\ dx, \mathrm{with}\ a = 0.63\ \mathrm{and}\ b = 2
\end{eqnarray}
In other words, the scaling exponent $\beta$ is at its minimum, when
\begin{eqnarray}
|\delta| & = & |Q| - 1
\end{eqnarray}
and at its maximum, when
\begin{eqnarray}
|Q| & = & \sqrt{|\delta|}
\end{eqnarray}

\begin{figure}[h]
\centering
\subfigure[][Frequency distribution of $\beta$]{
\includegraphics[width=0.43\textwidth]{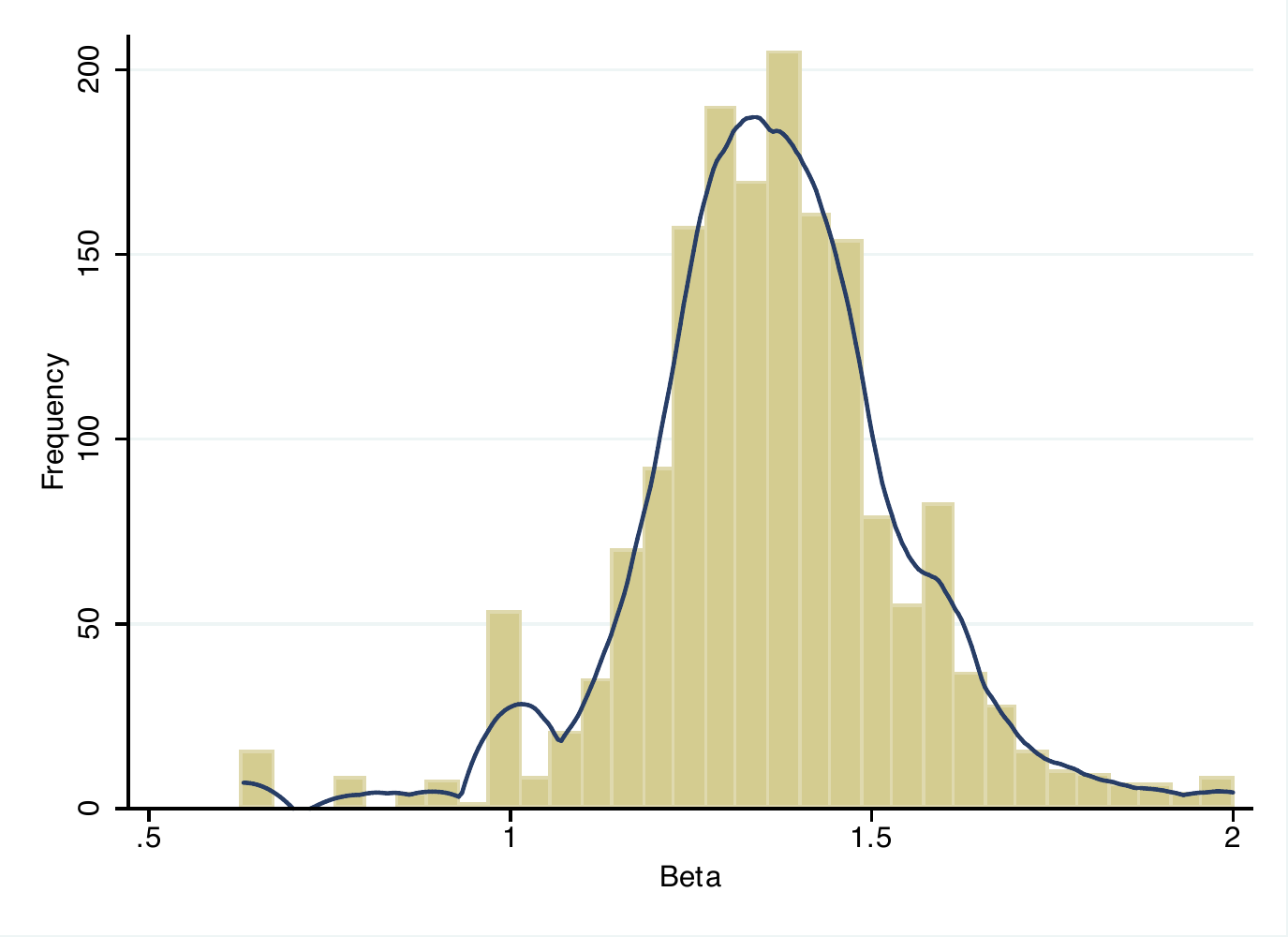}
\label{fig3A.fig}
}
\hspace{10mm}
\subfigure[][Evolution of $\beta$ vs. $|Q|$]{
\includegraphics[width=0.43\textwidth]{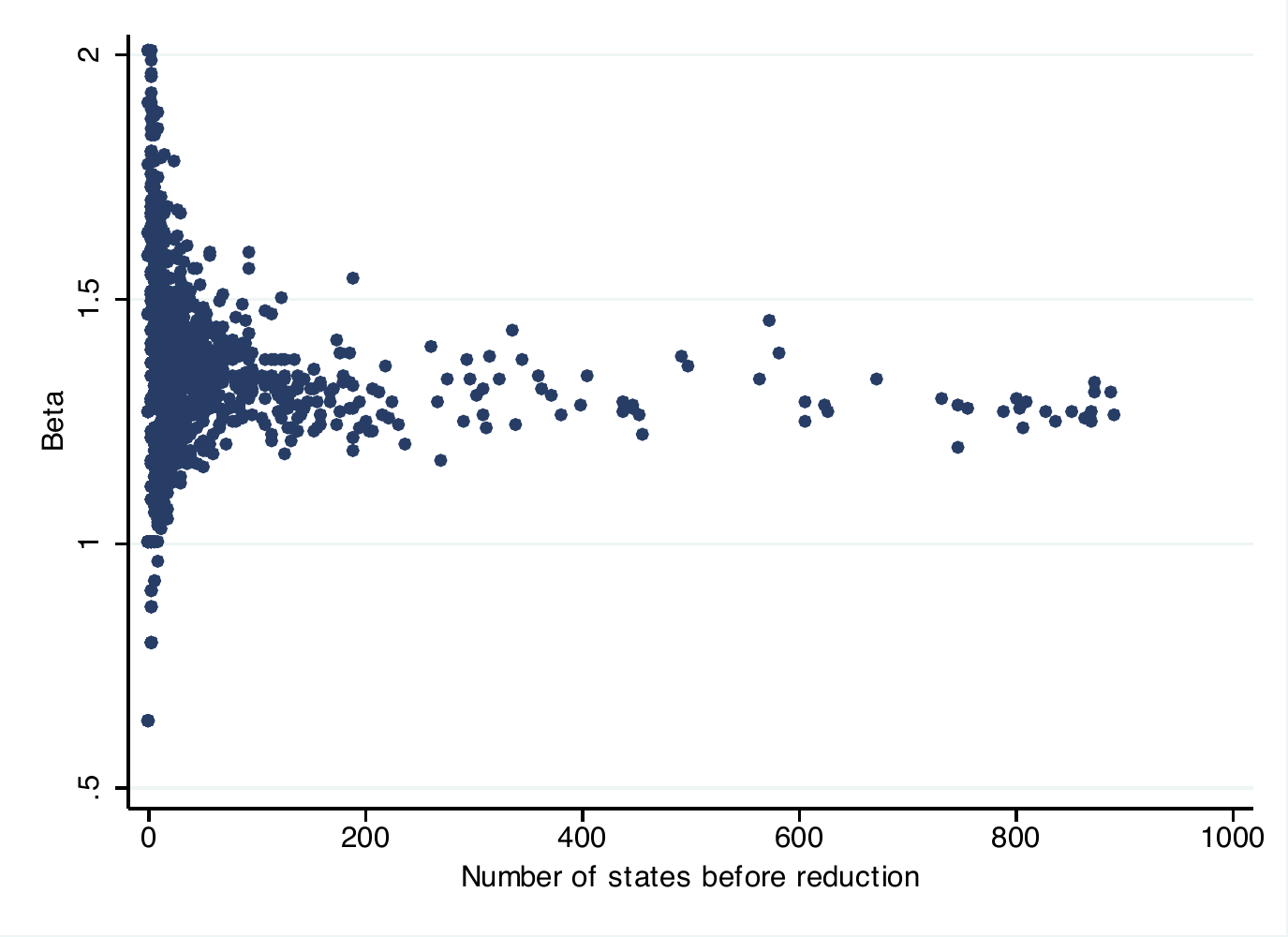}
\label{fig3B.fig}
}
\caption{The frequency distribution and evolution of the scaling exponent $\beta$.}
\label{betas.fig}
\end{figure}

The values of $\beta$ show remarkable similarity to those found in real-world software systems \cite{valverde:03a,vasa:05a}. The observed frequency distribution of $\beta$ for our experimental data set is shown in Figure~\ref{fig3A.fig}. The distribution of $\beta$ approximates a normal distribution with a mean value $\mu_\beta = 1.36$ and a standard deviation $\sigma_\beta = 0.19$.

As a system matures, with respect to a growing number of states, the scaling exponent $\beta$ converges towards the mean (cf. Figure~\ref{fig3B.fig}). Our experiments also confirm an observation made by Vasa \etal that the scaling exponents for real-world software systems plateaux at a system-specific value as the software systems mature \cite{vasa:05a}.

The Gini coefficient is a well-established measure to quantify the inequality of income distributions in moderns societies \cite{hdr:07a} that we have previously applied in the analysis of evolving software systems \cite{vasa:09a}. The Gini coefficient is a number between $0$ and $1$, where $0$ denotes a \emph{perfect equality} (\eg, every state in the system possesses the same number of outgoing transitions) and $1$ signifies a \emph{perfect inequality} (\eg, all states except one have no incoming transitions). The Gini coefficient is an entropic inequality measure. If its value is closer to $1$, then centralization of behavior in the system is greater with fewer states contributing to the information entropy \cite{shannon:01a} of the automaton. In other words, there are ``hubs" present in the automaton that centralize behavioral options and, therefore, yield a higher level of abstraction. With respect to partition refinement this means that the algorithm is more likely to succeed for an automaton that contains structural artifacts with Gini coefficients closer to $1$.

For a population with values $x_i$, $1 \le i \le n$, that are indexed in non-decreasing order ($x_i \le x_{i+1}$), the Gini coefficient is
\begin{eqnarray}
G & = & \frac{\Sigma^n_{i=1}(2i - n  - 1)x_i}{n\Sigma^n_{i=1}x_i}
\end{eqnarray}

\begin{figure}[h]
\centering
\subfigure[][Frequency distribution of $G_{IN}$]{
\includegraphics[width=0.45\textwidth]{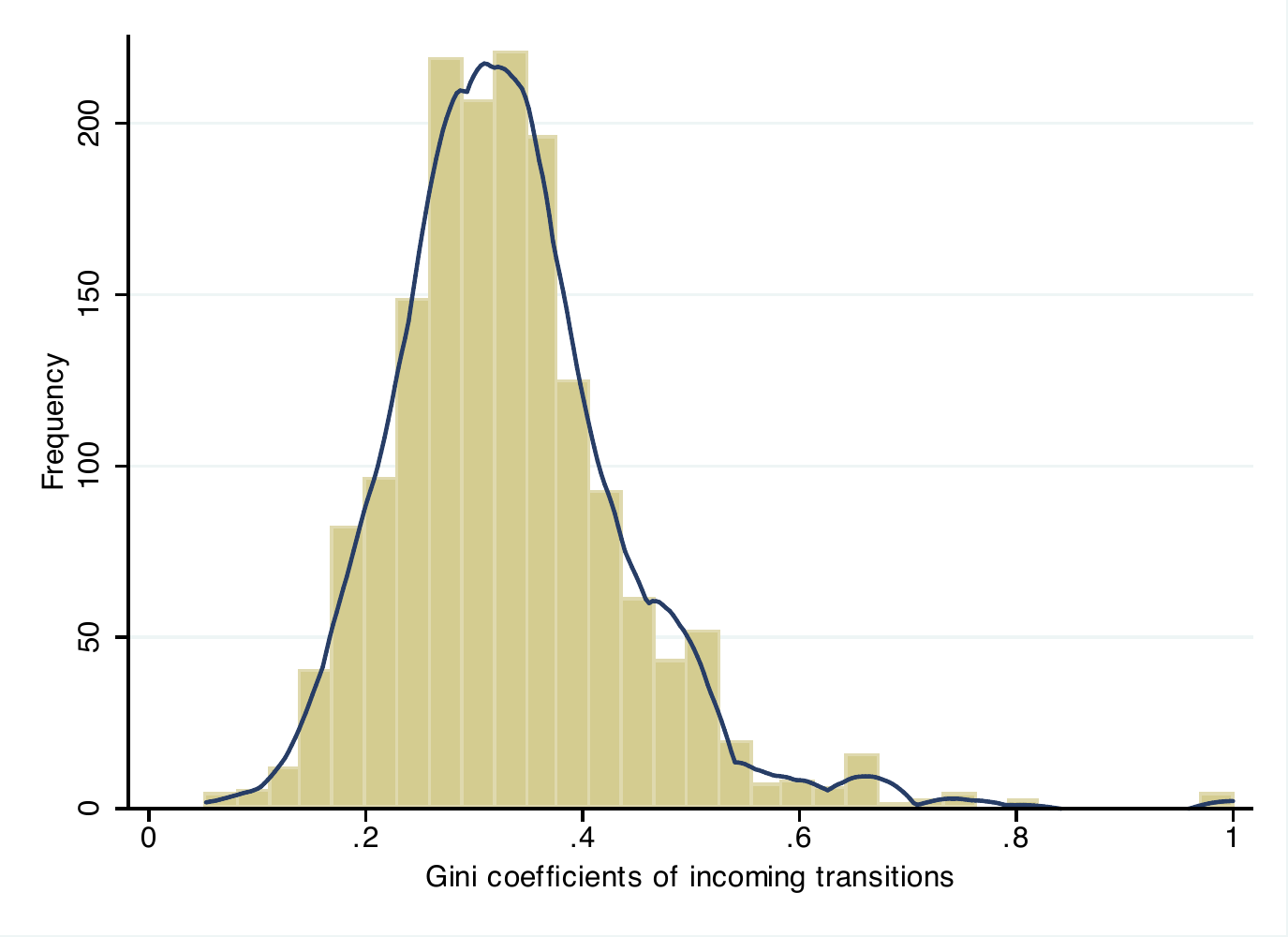}
\label{fig4A.fig}
}
\hspace{10mm}
\subfigure[][Frequency distribution of $G_{OUT}$]{
\includegraphics[width=0.45\textwidth]{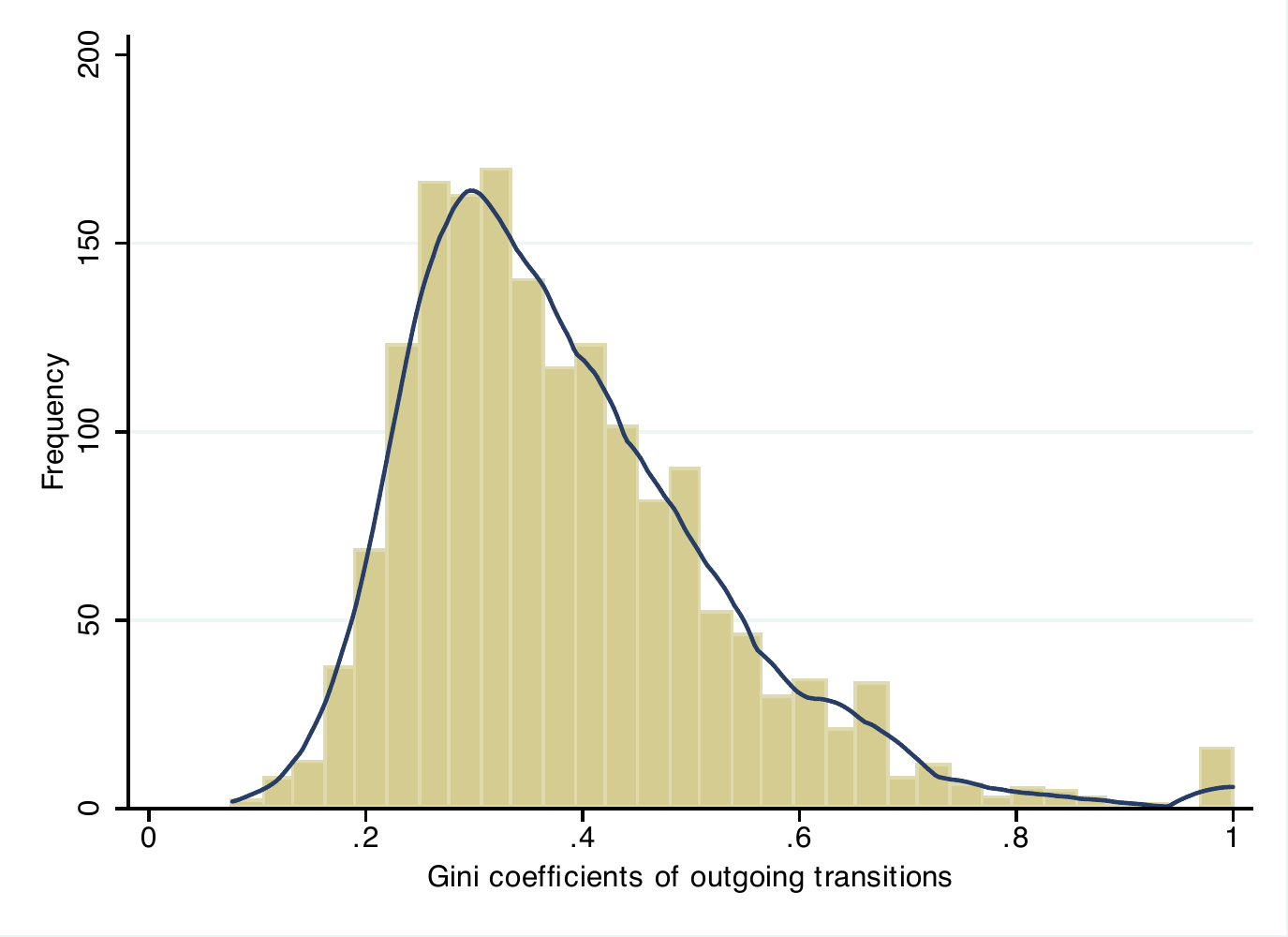}
\label{fig4B.fig}
}
\caption{The frequency distribution of Gini coefficients for incoming and outgoing transitions.}
\label{ginis.fig}
\end{figure}

We use $G_{IN}$ and $G_{OUT}$ to denote the Gini coefficient of incoming transitions and outgoing transitions, respectively. Observed value ranges of the Gini coefficients for our experimental automata are shown in Figure~\ref{ginis.fig}. The Gini coefficients of incoming transitions $G_{IN}$ (cf. Figure~\ref{fig4A.fig}) follow closely, though not perfectly, a normal distribution with mean value $\mu_{G_{IN}} = 0.34$ and a standard deviation $\sigma_{G_{IN}} = 0.11$. In order words, the distribution of incoming transitions in an automaton appears to be more likely independent of the behavior being modeled by the automaton.

\begin{figure}[t]
\centering
\subfigure[][5-State Synchronization Clique]{
\includegraphics[width=0.40\textwidth]{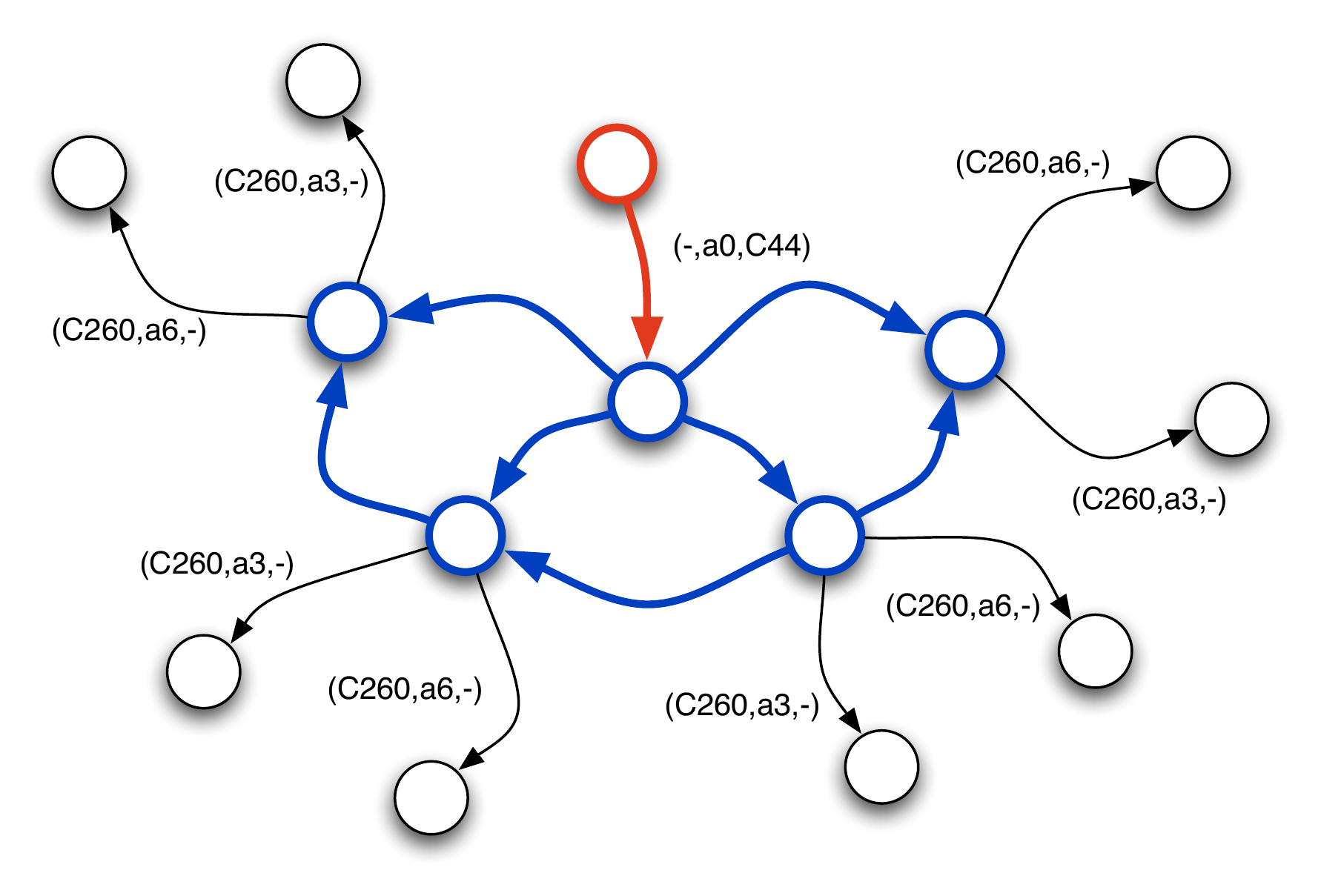}
\label{fig5A.fig}
}
\hspace{10mm}
\subfigure[][Emergent Hub]{
\includegraphics[width=0.40\textwidth]{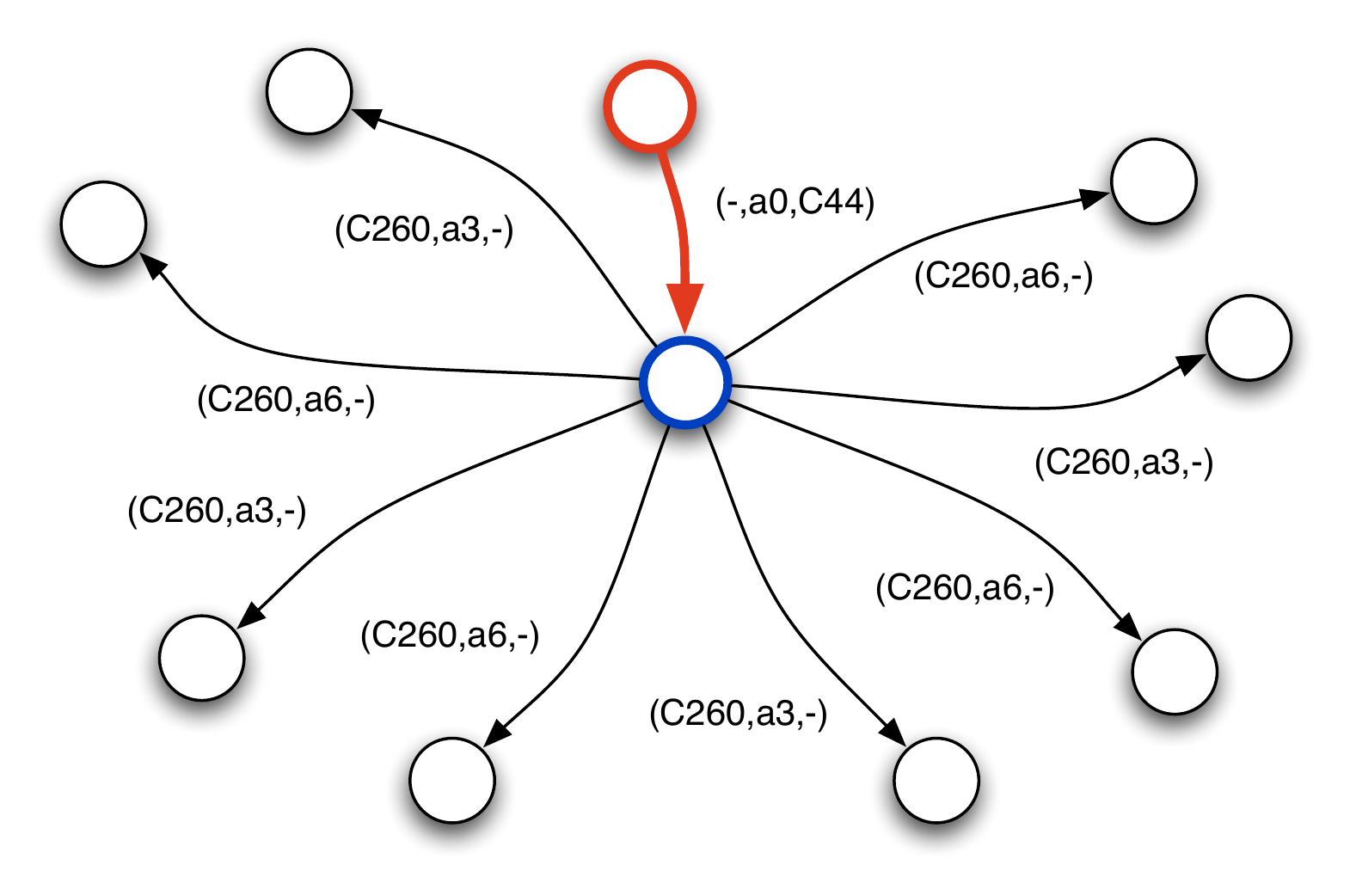}
\label{fig5B.fig}
}
\caption{Community structures in \emph{Component Interaction Automata}.}
\label{community.fig}
\end{figure}

In contrast, the values of the Gini coefficients for outgoing transitions $G_{OUT}$ (cf. Figure~\ref{fig4B.fig}) deviate significantly from a normal distribution. Even though the mean value $\mu_{G_{OUT}} = 0.38$ and the standard deviation $\sigma_{G_{OUT}} = 0.15$ are not very different from their respective $G_{IN}$ values, the values of $G_{OUT}$ exhibit a distinct positive skew with a fat tail of higher Gini coefficients. Skewed distributions emerge when parameters have multiplicative effects \cite{limpert:01} and certain factors prevail more than others. One such factor is \emph{preferential attachment} \cite{barabasi:99a} that favors behaviorally-rich states. However, there is another aspect to the concentration of outgoing transitions that is born of the partition refinement process itself. Consider Figure~\ref{fig5A.fig}, which depicts a fragment of the behavior of automaton $C260C44$. At the center of this fragment we find a set of states (marked in bold blue) that form a ``synchronization clique"{\footnote{We have omitted the labels to enhance readability.}} \cite{lumpe:10b}. A synchronization clique appears when states that are solely connected by internal component synchronizations become joined in one equivalence class by partition refinement. This is very typical for \emph{Component Interaction Automata} specifications and gives rise to significant state space reduction ratios \cite{lumpe:10b}. Figure~\ref{fig5B.fig} illustrates the corresponding effect of partition refinement. Where there were five states before, we find just one now that concentrates on itself a large number of outgoing transitions. In order words, we witness an ``emergent hub" in the system that acts as a focal point for previously disjoint behavioral choices. The fragment in Figure~\ref{fig5A.fig} has a Gini coefficient for outgoing transitions of $G_{OUT} = 0.67$, whereas the fragment in Figure~\ref{fig5B.fig} has $G_{OUT} = 0.89$. A higher $G_{OUT}$ is more likely to enable successful partition refinement than a smaller one.

% Regression.tex

\section{Regression Analysis}
\label{regression.sec}

\emph{Do the size and the structure of a Component Interaction Automata specification influence the probability for success of partition refinement?} In order to answer this question we conducted a number of \emph{regression analyses} \cite{agresti:02a,boslaugh:08a} to determine whether an automaton's size, structure, or both impact the actual outcome of partition refinement and, if so, how.  

In statistics, regression analysis provides a means to study possible relationships between variables. Regression analysis involves constructing models with one or more explanatory variables, $X_i, 1 \le i \le n$,  and a response variable $Y$. For the analysis of partition refinement of \emph{Component Interaction Automata} specifications we use a special form of analysis, called \emph{logistic regression} \cite{agresti:02a,boslaugh:08a}, that is applicable when the response variable is a \emph{dichotomy}. A logistic model can be described formally as follows. Let $\pi(x)=Pr(Y=1|X=x)=1 - Pr(Y=0,X=x)$ be the \emph{hypothesized proportion} of an expected value $x$ within a population $\pi$. The corresponding \emph{logistic regression model} \cite{agresti:02a} is
\begin{eqnarray}
\pi(x) & = & \frac{e^{(a + bx)}}{1 + e^{(a + bx)}}
\end{eqnarray}
where $a$ is called the \emph{intercept} and $b$ is called the \emph{regression coefficient}. 

The response variable $Y$ is categorical, where $1$ represents ``success" and $0$ denotes ``failure." We define two questions and store the corresponding answer in the associated response variables for our analysis: \emph{``Does partition refinement succeed?"} and \emph{``Does partition refinement require more than 5 minutes to converge?"}

The explanatory variable $X$ can be either numerical or categorical. For the analysis of partition refinement we use numerical variables. In particular, for a given automaton $\mathcal{C} = (Q, \mathit{Act}, \delta, I, H)$ we construct four different logistic models with the explanatory variable $X$ being either $|Q|$, the size of the automaton in terms of number of states, $\beta$, the power-scaling relationship between the number of states and the number of transitions, $G_{IN}$, the concentration of incoming transitions, or $G_{OUT}$, the concentration of outgoing transitions. 

\subsection{Experiment Setup and Analysis Approach}

We selected two random samples of \emph{Component Interaction Automata} specifications from a pool of 5,849 machine-generated candidates. The first set contained 840 automata for which partition refinement succeeded, whereas the second set, of equal size, comprised automata where partition refinement failed. In order to guarantee a meaningful comparison, we ensured that the automata in both sample sets have similar properties. The average size in terms of number of states is approximately 46, with 2 being the minimum and 892 being the maximum number of states in an automaton for both sets. Similarly, the reduction time for both sets of automata spans a range from 6 milliseconds to 2 hours.

We applied \emph{Maximum Likelihood Estimation} (MLE) to construct our logistic models and to determine the corresponding regression parameters $a$ and $b$. To verify that there exists a dependency between the explanatory variable $X$ and the response variable $Y$, we checked for each constructed model whether the likelihood ratio $\chi^2$ is significant at 1 degree of freedom (\emph{df.}) within a 95\% confidence interval (\emph{cf.}). If a logistic model shows a statistically significant relationship, then we use it to derive the probability of success for the full range of values of the explanatory variable and analyze how the probability of success changes with the value of the explanatory variable. 

In addition, to qualify the strength of the relationship between the explanatory variable and the response variable, we also compute two further measures: \emph{sensitivity} and \emph{specificity}. The former captures the probability of detecting a success of partition refinement when an actual reduction has occurred. The latter reflects the probability of detecting failure when partition refinement has indeed failed. Sensitivity and specificity provide guarantees that a constructed regression model is effective at detecting equally well both success and failure and is, in fact, better than pure random guessing \cite{agresti:02a}.

Ideally, the values for sensitivity and specificity should be as close to 100\% as possible. In this case the model will correctly classify all successes and failures using just the explanatory variable. However, in practice models are never perfect. To further improve on the predictive power of the explanatory variable, we require the values of sensitivity and specificity to exceed 50\%  \cite{agresti:02a} by a comfortable margin. Due to the composition of our sample set, partition refinement succeeds, by default, in 50\% of the cases. So, if for a given model the values of sensitivity and specificity are just around 50\%, then the model only confirms the threshold already being embedded in our sample data set. Hence, only if we achieve values for sensitivity and specificity greater than 50\% will the corresponding explanatory variable become a suitable predictor for the success or failure of partition refinement.

\subsection{Impact of Size on Partition Refinement}

\begin{figure}[t]
\centering
\subfigure[][Partition refinement requires more than 5mins]{
\includegraphics[width=0.45\textwidth]{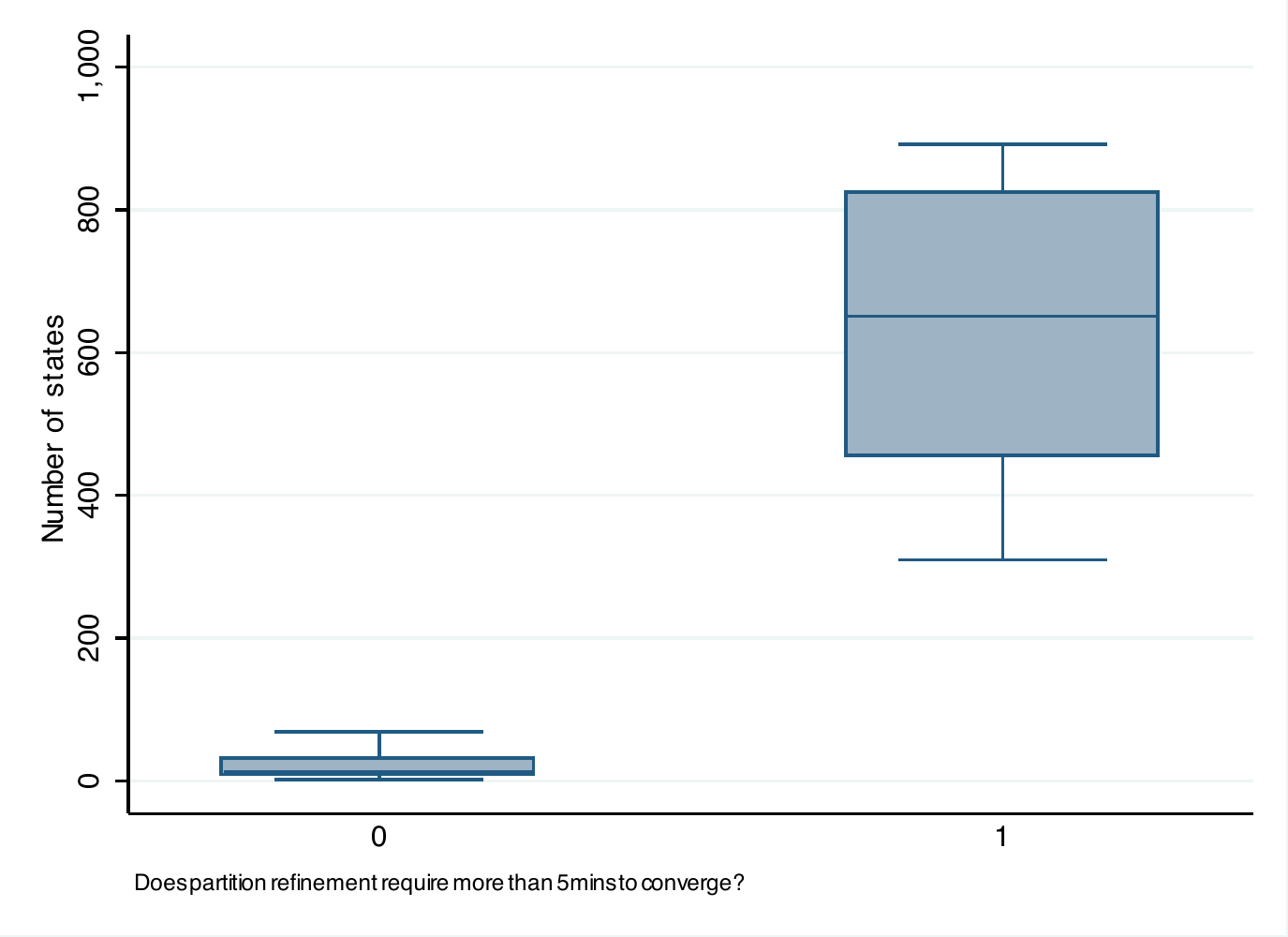}
\label{fig6A.fig}
}
\hspace{10mm}
\subfigure[][The influence of $|Q|$ on running time]{
\includegraphics[width=0.45\textwidth]{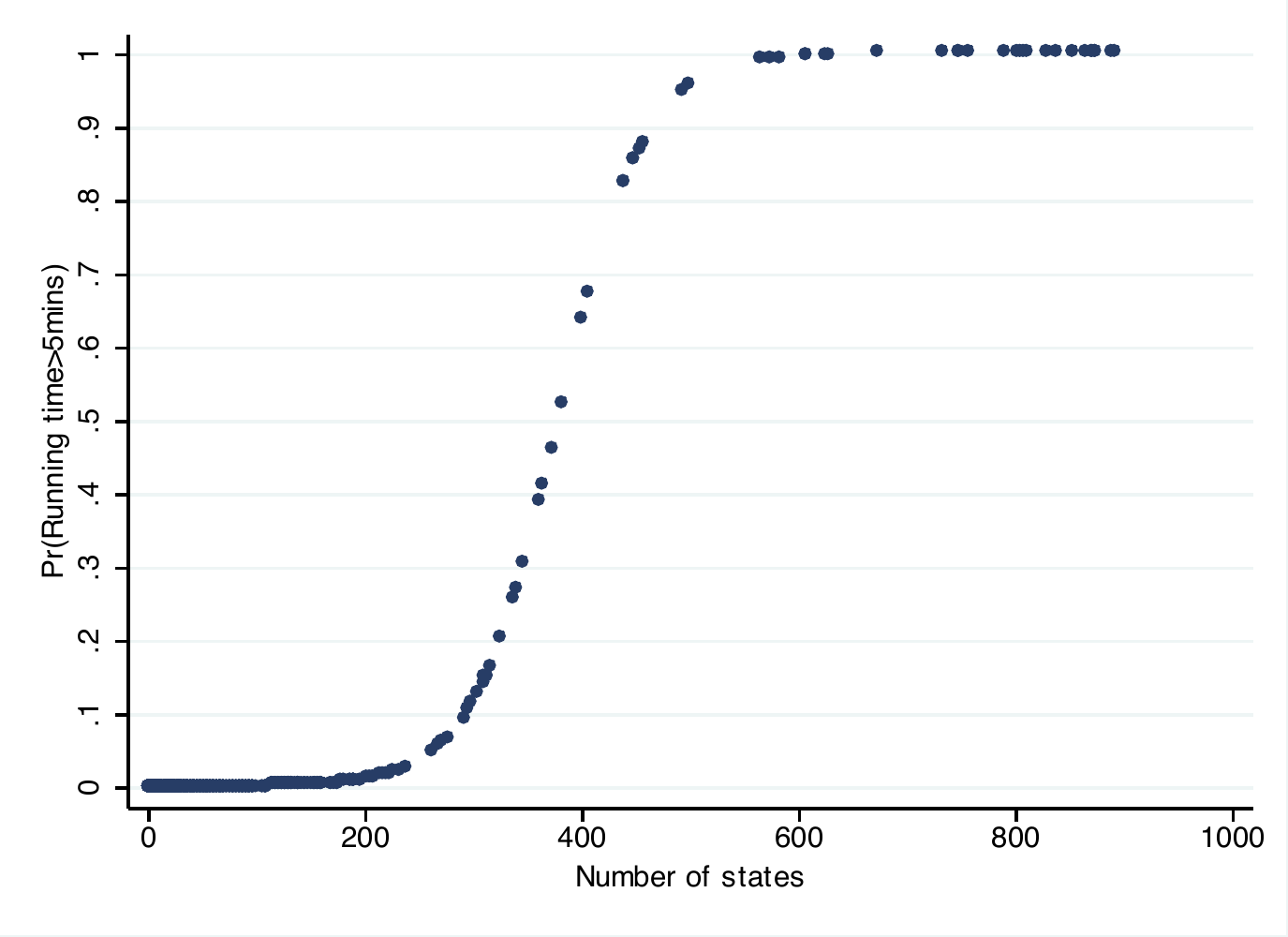}
\label{fig6B.fig}
}
\caption{A logistic model for the prediction of running time of partition refinement.}
\label{statesfivemins.fig}
\end{figure}

The size of an automaton has a direct impact on the running time of partition refinement of \emph{Component Interaction Automata} specifications, as it is known to have exponential time and space complexity \cite{lumpe:08a}. Using the upper limit of 5 minutes, an observed suitable threshold, we can construct a logistic model that reliably predicts whether partition refinement will require more than 5 minutes based on the size, in terms of number of states $|Q|$, of an automaton. Figure~\ref{statesfivemins.fig} illustrates the model-specific values. We can expect partition refinement for systems with less than 200 states to always converge in less than 5 minutes. However, we have also observed cases in which partition refinement converged for automata with up to 450 states within that limit. But above 450 states, partition refinement is likely to require more than 5 minutes, as indicated in Figure~\ref{fig6A.fig}. 

The corresponding logistic model (cf. Figure~\ref{fig6B.fig}) confirms these estimates. In fact, the number of states of an automaton $|Q|$ is a suitable predictor for the running time of partition refinement (\ie, \mbox{$\chi^2$ = $342.68$,} 1 \emph{df.} at 95\% \emph{cf.} with p-value of $0.0001$).  Sensitivity is at 87\% and specificity at 99.8\%. The model asserts that if an automaton reaches more than 385 states, partition refinement will require, with a probability of $>0.5$, more than 5 minutes to converge. Moreover, the probability of partition refinement to require more than 5 minutes is $1$ for automata with more than 570 states. 

However, size is not a suitable predictor for the success of partition refinement. We are unable to construct a corresponding model ($\chi^2$ = $0.01$, 1 \emph{df.} at 95\% \emph{cf.} with p-value of $0.928$). Based on our analysis we find that the size of an automaton and success of partition refinement are independent variables. Partition refinement may succeed or fail independent of the actual size of the automaton. However, this does not mean that we can ignore size when considering the success of partition refinement. There appears to be a natural resistance to successful partition refinement when automata increase in size.

\subsection{Impact of Structure on Partition Refinement}

The size of an automaton does not yield a good predictor for the success of partition refinement. But what is the impact of structure on partition refinement, in particular with respect to a successful state space reduction?

\begin{figure}[t]
\centering
\subfigure[][Partition refinement success]{
\includegraphics[width=0.45\textwidth]{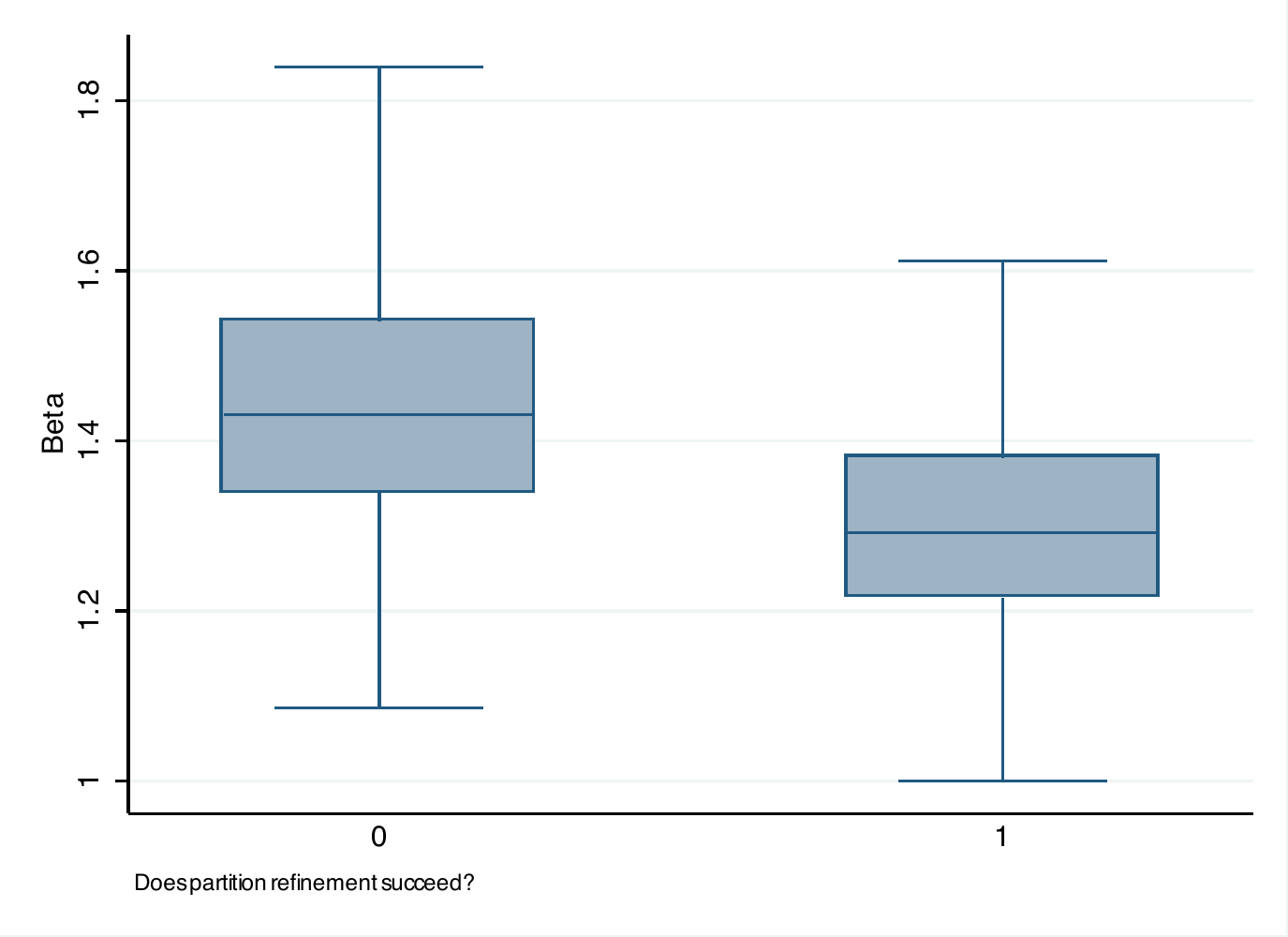}
\label{fig7A.fig}
}
\hspace{10mm}
\subfigure[][The influence of $\beta$]{
\includegraphics[width=0.45\textwidth]{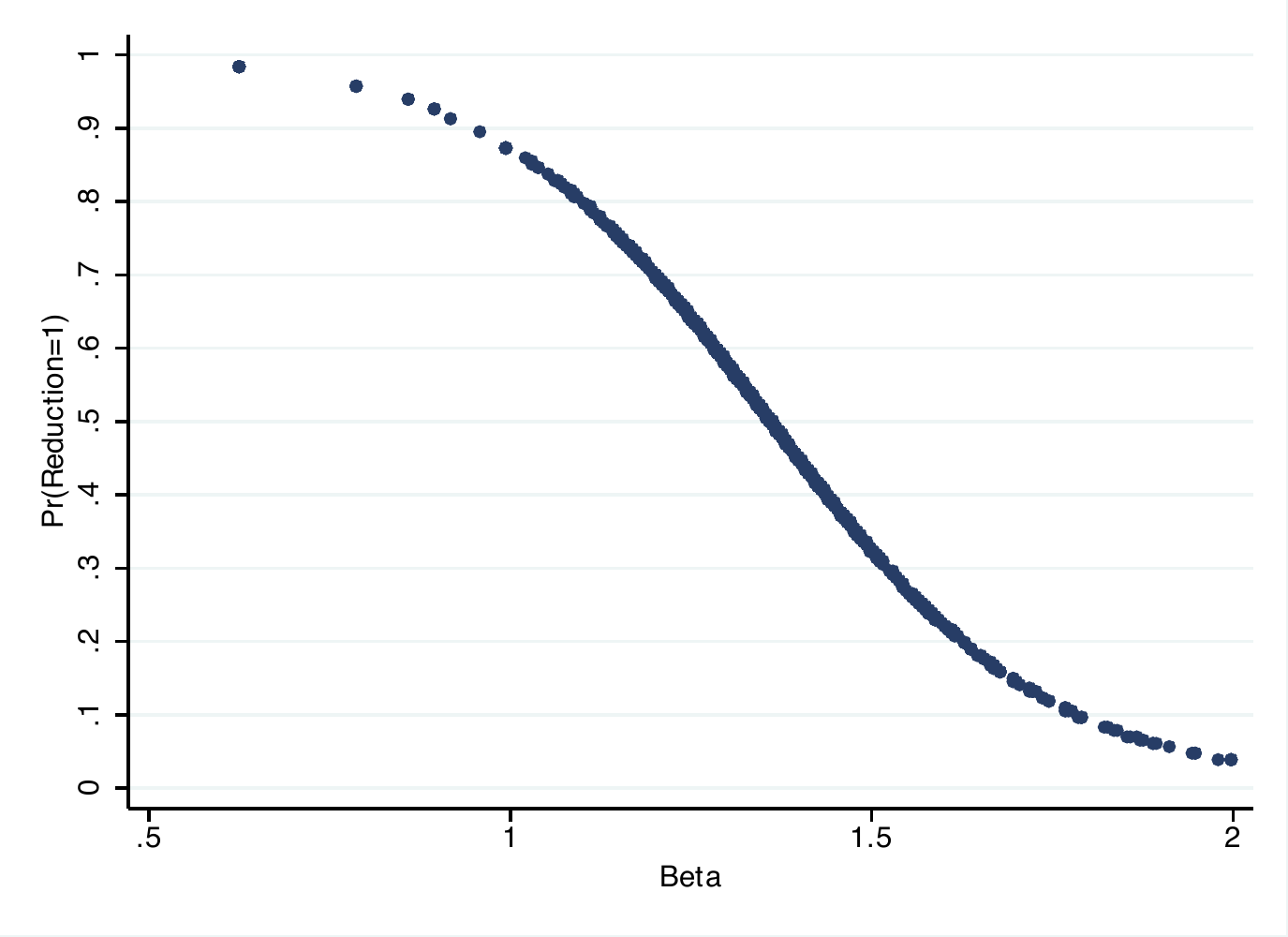}
\label{fig7B.fig}
}
\caption{A logistic model for the prediction of success of partition refinement based on $\beta$.}
\label{betareduction.fig}
\end{figure}

We have selected three topological attributes: $\beta$, $G_{IN}$, and $G_{OUT}$. Does the ratio between states and transitions in terms of the power-scaling relationship $|\delta| \sim |Q|^\beta$ provide us with a predictor for the success of partition refinement? The answer is yes. Consider Figure~\ref{betareduction.fig} that presents our observations for the scaling exponent $\beta$. There is a significant difference in the distribution of the successes and failures of partition refinement (cf. Figure~\ref{fig7A.fig}). We find that automata with higher $\beta$ values are less likely candidates for successful partition refinement than those with smaller $\beta$ values. The mean value of $\beta$ for success is $\mu^1_\beta = 1.29$, whereas that for failure is $\mu^0_\beta = 1.43$. Based on these observations it appears that partition refinement becomes more likely to succeed $\frac{1}{2}\sigma_\beta$ below the mean value $\mu_\beta$ and the chances for success diminish increasingly $\frac{1}{2}\sigma_\beta$ above of the mean value $\mu_\beta$.

The logistic model (cf. Figure~\ref{fig7B.fig}) confirms this. The scaling factor $\beta$ yields a good predictor for the success of partition refinement (\ie, $\chi^2$ = $279.61$, 1 \emph{df.} at 95\% \emph{cf.} with p-value of $0.0001$).
Both the sensitivity and the specificity are at 71\%, suggesting that the model is strong in detecting the probability of success (sensitivity) and the probability of failure (specificity), respectively. The model corroborates the view that if the scaling exponent $\beta$ increases the probability for a successful partition refinement decreases.

There is another intriguing aspect to this model. As the \emph{Component Interaction Automata} specifications for real-world systems mature the scaling exponents plateaux at the mean value $\mu_\beta = 1.36$. This is exactly the value at which the model predicts the probability for the success of partition refinement to be $0.5$. As a result, this suggests that one in two \emph{Component Interaction Automata} specification for real-world software systems can be reduced by partition refinement. In fact, the odds are slightly in favor of success for partition refinement of real-world software system specifications (cf. Grindstead and Snell's anecdote of \emph{Chevalier de M{\'er\'e}'s rolling dice bet} \cite{change:06a}). The probability for success is actually somewhat above $0.5$, as rounding of $\mu_\beta$ pushes its value up.

The concentration of incoming transitions $G_{IN}$ fails to serve as a predictor for the success of partition refinement (\ie, $\chi^2$ = $1.76$, 1 \emph{df.} at 95\% \emph{cf.} with p-value of $0.1847$). However, this does not come as a surprise. The Gini coefficient for incoming transitions appears to be independent of the modeled behavior. As far as $G_{IN}$ is concerned, the success of partition refinement cannot be predicted. From the point of partition refinement, it matters more how many behavioral variants a state can produce than how many behavioral variants a state depends on.

\begin{figure}[t]
\centering
\subfigure[][Partition refinement is successful?]{
\includegraphics[width=0.45\textwidth]{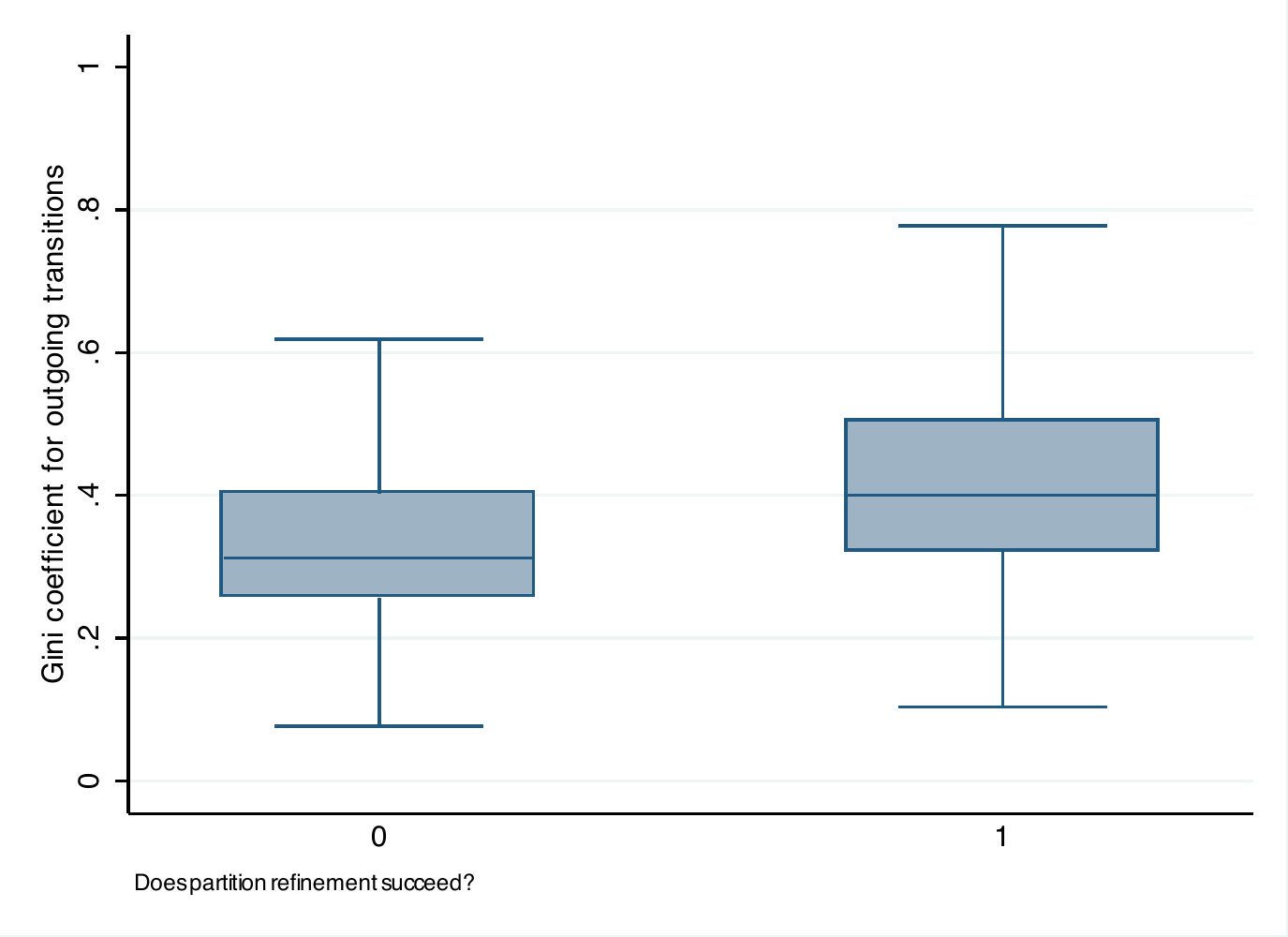}
\label{fig8A.fig}
}
\hspace{10mm}
\subfigure[][The influence of $G_{OUT}$.]{
\includegraphics[width=0.45\textwidth]{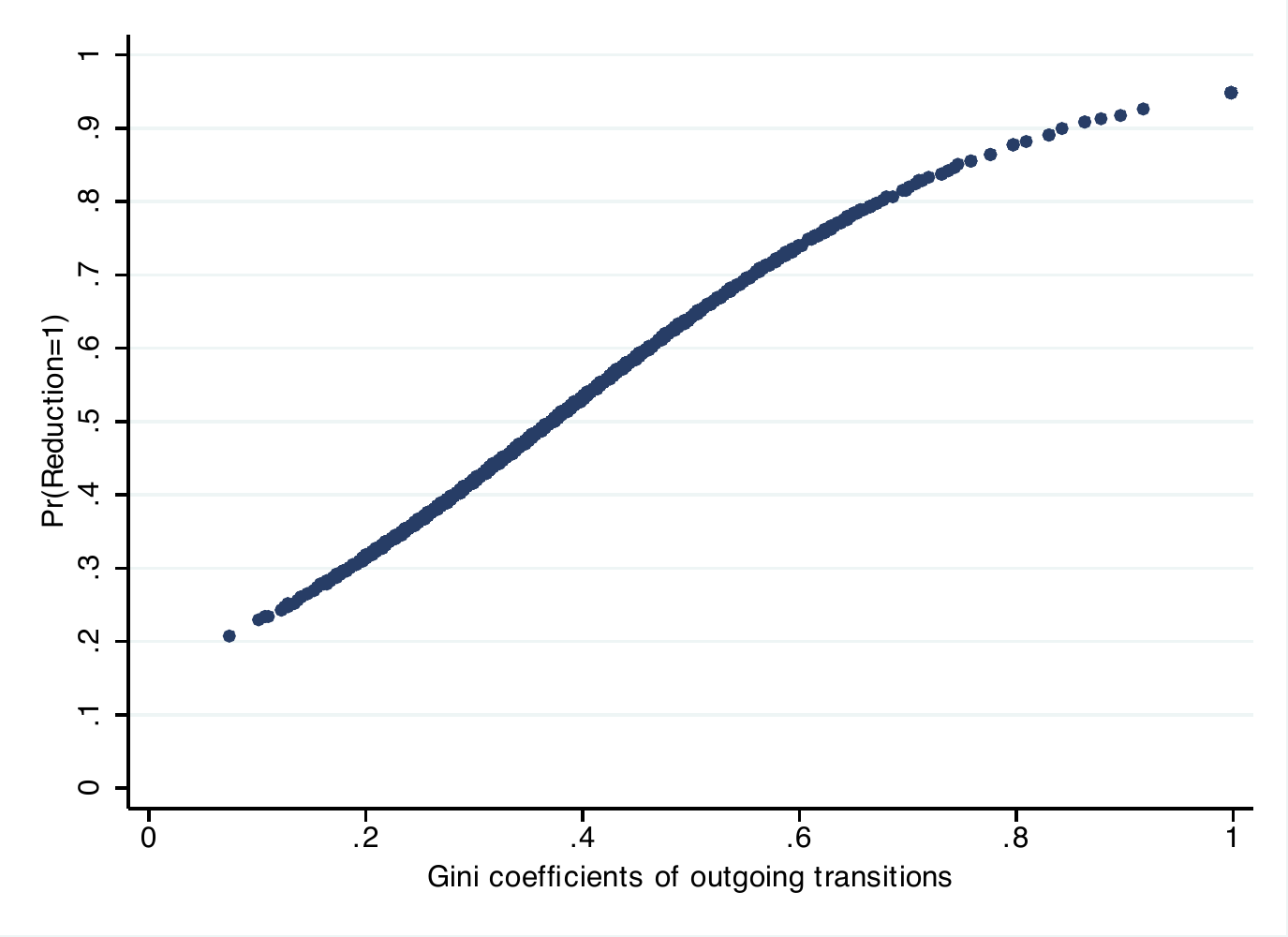}
\label{fig8B.fig}
}
\caption{A logistic model for the prediction of success of partition refinement based on $G_{OUT}$.}
\label{ginioutreduction.fig}
\end{figure}

Finally, we explored the concentration of outgoing transitions $G_{OUT}$. Unlike $G_{IN}$, $G_{OUT}$ furnishes us with a suitable predictor for the success of partition refinement. Even though the margin is small, it is sufficiently significant to provide us with a discriminator for the success of partition refinement. Partition refinement is more likely to fail if the value of $G_{OUT}$ moves towards $0.34$. In contrast, if the value of $G_{OUT}$ moves closer to $0.43$, partition refinement is  more likely to succeed (cf. Figure~\ref{fig8A.fig}). There is a narrow margin, $\pm$4\%, that determines the success or failure of partition refinement. This value is of specific significance, as it corresponds exactly to the threshold defined by Vasa \etal \cite{vasa:09a} for the identification of major shifts in evolving software systems. In other words, a deviation from the mean value $\mu_{G_{OUT}}$ by more than 4\% significantly influences the success of partition refinement.

We can construct a logistic model using $G_{OUT}$ (\ie, $\chi^2$ = $148.63$, 1 \emph{df.} at 95\% \emph{cf.} with p-value of $0.0001$). However, sensitivity and specificity are not as strong as in the case of the model for $\beta$. In fact, the model is stronger at detecting failure (specificity at 69\%) and weaker at detecting  success (sensitivity at 57\%) of partition refinement. Nevertheless, it is still more reliable than a simple guess.

The models for $\beta$ and $G_{OUT}$ offer quite opposite directions. Partition refinement is expected to more likely succeed if the automaton exhibits a low $\beta$ value and a high $G_{OUT}$ value. In other words, a large amount of behavioral choices in selected states can assist partition refinement to produce a smaller state space. However, the number of these behavioral choices has to be balanced with the total amount of choice points in an automaton, as indicated by the model for $\beta$. 

% Conc.tex

\section{Conclusion}
\label{conc.sec}

\emph{Component Interaction Automata} provide a fitting technique to capture and analyze the temporal facets of hierarchical-structured component-based systems. It is, however, in the nature of automata-based approaches that the respective specifications suffer from a combinatorial state explosion problem. For this reason, an effective use of \emph{Component Interaction Automata} for the specifications and analysis of real-world software systems may become difficult, if not impossible, due to the underlying complexity of the systems being modeled. We, therefore, seek to find suitable abstraction methods that can help us to cope with the state explosion problem.

Partition refinement through weak bisimulation can alleviate the impact of state explosion, but this technique too exhibits exponential time and space complexity \cite{lumpe:08a}. Worse, success is erratic. To better understand why this abstraction technique succeeds in some cases and fails in others, we have conducted an empirical study on 1,680 \emph{Component Interaction Automata} specifications and constructed several logistic regression models that can explain the observed performance of partition refinement. We learn that structure, not size, has a bigger impact on the success of partition refinement. However, we cannot completely dismiss size as a contributing factor to the success of partition refinement. Eventually, the partition refinement algorithm will succumb to the size of an automaton. Even though the topology of an automaton can positively influence of the outcome of partition refinement, we must not neglect size altogether as it affects the running time of partition refinement. 

\begin{figure}[t]
\centering
\subfigure[][Likelyhood of achieving 50\% state space reduction]{
\includegraphics[width=0.45\textwidth]{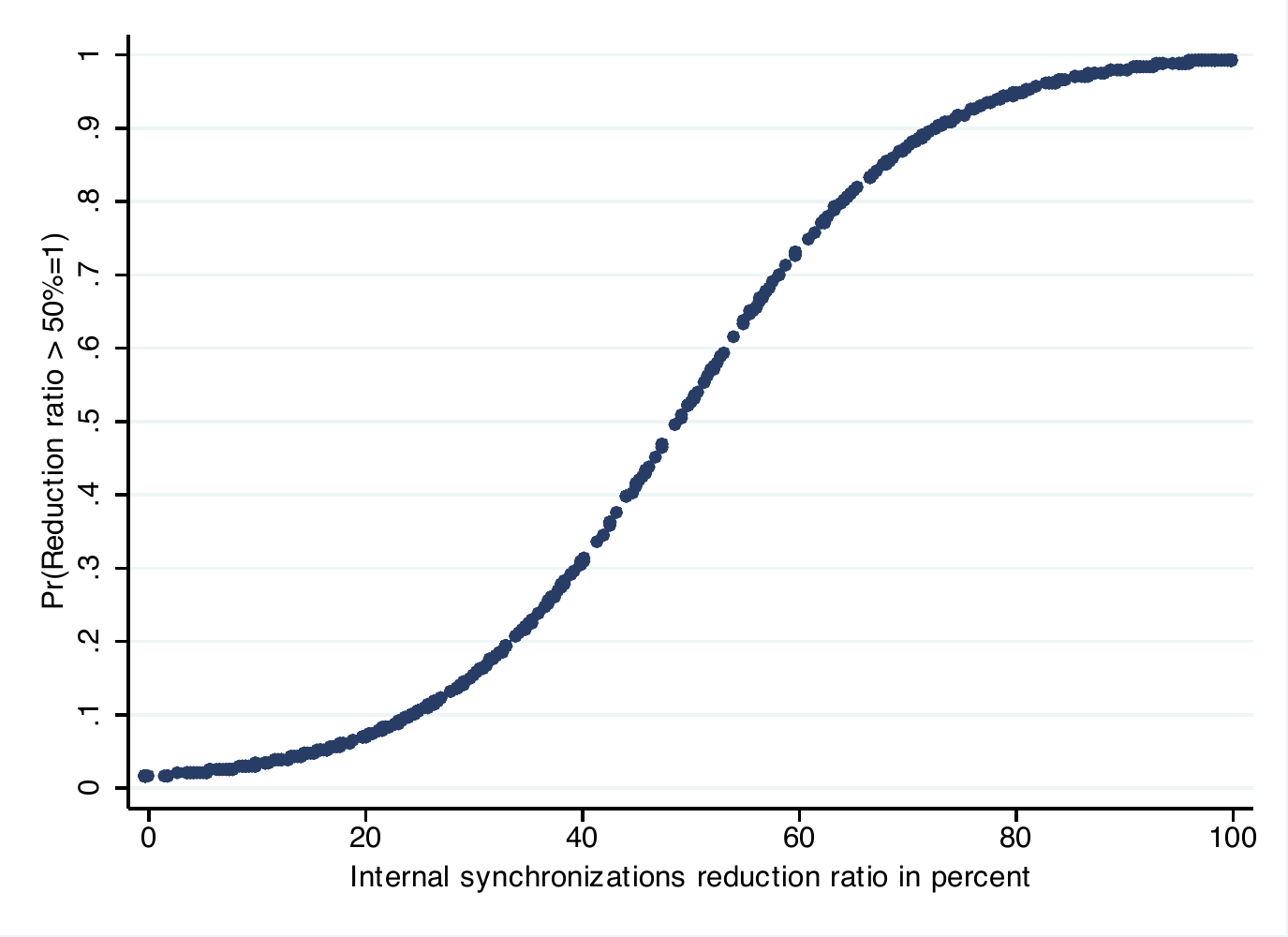}
\label{fig9A.fig}
}
\hspace{10mm}
\subfigure[][Likelyhood of achieving 75\% state space reduction]{
\includegraphics[width=0.45\textwidth]{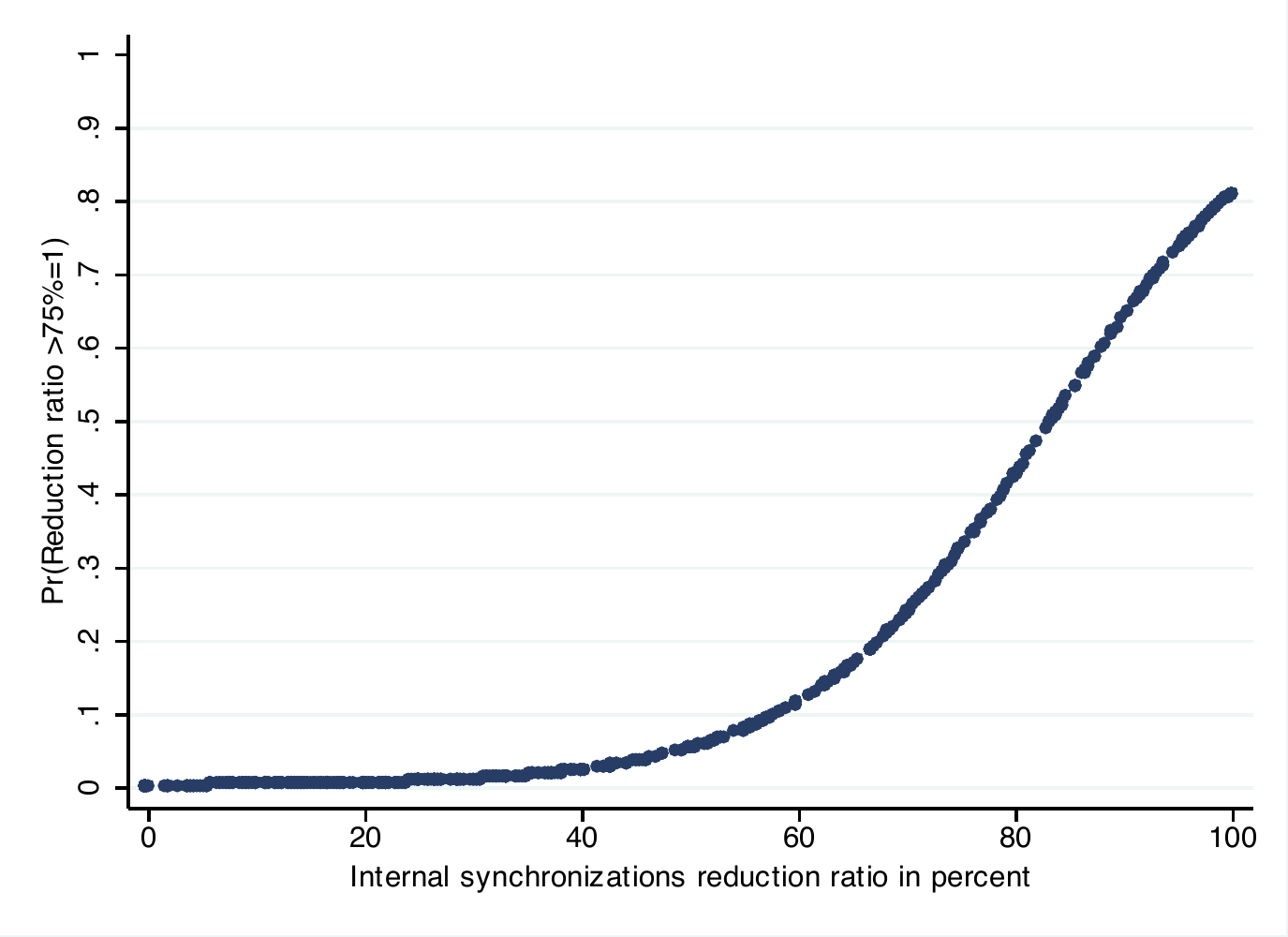}
\label{fig9B.fig}
}
\caption{The effectiveness of partition refinement.}
\label{reductionratio.fig}
\end{figure}

Partition refinement can achieve excellent results and yield strong state space reduction ratios (cf. Figure~\ref{reductionratio.fig}). However, the results depend on the presence of ``synchronization cliques", community structures that partition refinement can eliminate in the refinement process (cf. Figure~\ref{community.fig}). These cliques have to be of sufficient size to have an impact. For example, we need to be able to remove 84\% or more internal synchronizations from an automaton in order to achieve an overall state space reduction ratio of 75\% or more (cf. Figure~\ref{fig9B.fig}). But again, the internal synchronizations must be of the right kind -- members of cliques. Simply having many internal synchronizations occurring in an automaton does not suffice -- they have to occur in the right structural artifacts. It is the structure, not size, that influences most the outcome of the refinement process.

The application of \emph{Component Interaction Automata} for the specification and analysis of component-based software systems is similar in character to a ``non-cooperative game" \cite{nash:50a}. There are competing forces at work that need to be balanced in order to achieve the desired outcome. Automata-based techniques can be used for the specification of real-world software systems, but the level of granularity, in terms of both structure and size, has to be chosen carefully to compensate for the inherent and inevitable associated state space explosion.

% end body

\bibliographystyle{eptcs}

\begin{thebibliography}{10}
\providecommand{\bibitemstart}[1]{\bibitem{#1}}
\providecommand{\bibitemend}{}
\providecommand{\bibliographystart}{}
\providecommand{\bibliographyend}{}
\providecommand{\url}[1]{\texttt{#1}}
\providecommand{\urlprefix}{Available at }
\providecommand{\bibinfo}[2]{#2}
\bibliographystart

\bibitemstart{agresti:02a}
\bibinfo{author}{Alan Agresti} (\bibinfo{year}{2002}):
  \emph{\bibinfo{title}{{Categorical Data Analysis}}}.
\newblock \bibinfo{publisher}{John Wiley \& Sons, Inc.},
  \bibinfo{address}{Hoboken, New Jersey}, \bibinfo{edition}{second} edition.
\bibitemend

\bibitemstart{alfaro:01a}
\bibinfo{author}{Luca de~Alfaro} \& \bibinfo{author}{Thomas~A. Henzinger}
  (\bibinfo{year}{2001}): \emph{\bibinfo{title}{{Interface Automata}}}.
\newblock In: \bibinfo{editor}{Volker Gruhn} \& \bibinfo{editor}{A~Min Tjoa},
  editors: {\sl \bibinfo{booktitle}{Proceedings ESEC/FSE 2001}},
  \bibinfo{publisher}{ACM Press}, \bibinfo{address}{Vienna, Austria}, pp.
  \bibinfo{pages}{109--120}.
\bibitemend

\bibitemstart{barabasi:99a}
\bibinfo{author}{Albert-L{\'a}szl{\'o} Barab{\'a}si} \&
  \bibinfo{author}{R{\'e}ka Albert} (\bibinfo{year}{1999}):
  \emph{\bibinfo{title}{{Emergence of Scaling in Random Networks}}}.
\newblock {\sl \bibinfo{journal}{Science}}
  \bibinfo{volume}{286}(\bibinfo{number}{5439}), pp. \bibinfo{pages}{509--512}.
\bibitemend

\bibitemstart{beek:03a}
\bibinfo{author}{Maurice~H. ter Beek}, \bibinfo{author}{Clarence~A. Ellis},
  \bibinfo{author}{Jetty Kleijn} \& \bibinfo{author}{Grzegorz Rozenberg}
  (\bibinfo{year}{2003}): \emph{\bibinfo{title}{{Synchronizations in Team
  Automata for Groupware Systems}}}.
\newblock {\sl \bibinfo{journal}{Computer Supported Cooperative Work}}
  \bibinfo{volume}{12}(\bibinfo{number}{1}), pp. \bibinfo{pages}{21--69}.
\bibitemend

\bibitemstart{beugnard:99}
\bibinfo{author}{Antoine Beugnard}, \bibinfo{author}{Jean-Marc
  J{\'e}z{\'e}quel}, \bibinfo{author}{No{\"e}l Plouzeau} \&
  \bibinfo{author}{Damien Watkins} (\bibinfo{year}{1999}):
  \emph{\bibinfo{title}{{Making Components Contract Aware}}}.
\newblock {\sl \bibinfo{journal}{IEEE Computer}}
  \bibinfo{volume}{32}(\bibinfo{number}{7}), pp. \bibinfo{pages}{38--45}.
\bibitemend

\bibitemstart{boslaugh:08a}
\bibinfo{author}{Sarah Boslaugh} \& \bibinfo{author}{Paul~Andrew Watters}
  (\bibinfo{year}{2008}): \emph{\bibinfo{title}{{Statistics in a Nutshell -- A
  Desktop Quick Reference}}}.
\newblock \bibinfo{publisher}{O'Reilly Media Inc.},
  \bibinfo{address}{Sebastopol, California}.
\bibitemend

\bibitemstart{brim:06a}
\bibinfo{author}{Lubo\v{s} Brim}, \bibinfo{author}{Ivana \v{C}ern\'{a}},
  \bibinfo{author}{Pavl\'{\i}na Va\v{r}ekov\'{a}} \& \bibinfo{author}{Barbora
  Zimmerova} (\bibinfo{year}{2006}):
  \emph{\bibinfo{title}{{Component-Interaction Automata as a
  Verification-Oriented Component-Based System Specification}}}.
\newblock {\sl \bibinfo{journal}{SIGSOFT Software Engineering Notes}}
  \bibinfo{volume}{31}(\bibinfo{number}{2}), pp. \bibinfo{pages}{1--8}.
\bibitemend

\bibitemstart{cerna:07a}
\bibinfo{author}{Ivana {\v{C}}ern{\'a}}, \bibinfo{author}{Pavl{\'i}na
  Va{\v{r}}ekov{\'a}} \& \bibinfo{author}{Barbora Zimmerova}
  (\bibinfo{year}{2007}): \emph{\bibinfo{title}{{Component Substitutability via
  Equivalencies of Component-Interaction Automata}}}.
\newblock {\sl \bibinfo{journal}{Electronic Notes in Theoretical Computer
  Science}} \bibinfo{volume}{182}, pp. \bibinfo{pages}{39--55}.
\bibitemend

\bibitemstart{fortunato:09a}
\bibinfo{author}{Santo Fortunato} (\bibinfo{year}{2009}):
  \emph{\bibinfo{title}{{Community detection in graphs}}}.
\newblock {\sl \bibinfo{journal}{Physics Reports}} , pp.
  \bibinfo{pages}{75--174}.
\bibitemend

\bibitemstart{gini:21a}
\bibinfo{author}{Corrado Gini} (\bibinfo{year}{1921}):
  \emph{\bibinfo{title}{{Measurement of Inequality of Incomes}}}.
\newblock {\sl \bibinfo{journal}{The Economic Journal}}
  \bibinfo{volume}{31}(\bibinfo{number}{121}), pp. \bibinfo{pages}{124--126}.
\bibitemend

\bibitemstart{hermanns:02a}
\bibinfo{author}{Holger Hermanns} (\bibinfo{year}{2002}):
  \emph{\bibinfo{title}{{Interactive Markov Chains: The Quest for Quantified
  Quality}}}.
\newblock \bibinfo{series}{LNCS 2428}. \bibinfo{publisher}{Springer},
  \bibinfo{address}{Heidelberg, Germany}.
\bibitemend

\bibitemstart{hopcroft:07a}
\bibinfo{author}{John~E. Hopcroft}, \bibinfo{author}{Rajeev Motwani} \&
  \bibinfo{author}{Jeffrey~D. Ullman} (\bibinfo{year}{2007}):
  \emph{\bibinfo{title}{{Automata Theory, Languages, and Computation}}}.
\newblock \bibinfo{publisher}{Pearson Education}, \bibinfo{edition}{3rd}
  edition.
\bibitemend

\bibitemstart{limpert:01}
\bibinfo{author}{Eckhard Limpert}, \bibinfo{author}{Werner~A. Stahel} \&
  \bibinfo{author}{Markus Abbt} (\bibinfo{year}{2001}):
  \emph{\bibinfo{title}{{Log-normal Distributions across the Sciences: Keys and
  Clues}}}.
\newblock {\sl \bibinfo{journal}{BioScience}}
  \bibinfo{volume}{51}(\bibinfo{number}{5}), pp. \bibinfo{pages}{341--352}.
\bibitemend

\bibitemstart{lumpe:99a}
\bibinfo{author}{Markus Lumpe} (\bibinfo{year}{1999}): \emph{\bibinfo{title}{{A
  $\pi$-Calculus Based Approach to Software Composition}}}.
\newblock \bibinfo{type}{Ph.D. thesis}, \bibinfo{school}{University of Bern,
  Institute of Computer Science and Applied Mathematics}.
\bibitemend

\bibitemstart{lumpe:10b}
\bibinfo{author}{Markus Lumpe} (\bibinfo{year}{2010}):
  \emph{\bibinfo{title}{{Action Prefixes: Reified Synchronization Paths in
  Minimal Component Interaction Automata}}}.
\newblock {\sl \bibinfo{journal}{Electronic Notes in Theoretical Computer
  Science}} \bibinfo{volume}{263}, pp. \bibinfo{pages}{179--195}.
\newblock \bibinfo{note}{Proceedings of the 6th International Workshop on
  Formal Aspects of Component Software (FACS 2009)}.
\bibitemend

\bibitemstart{lumpe:08a}
\bibinfo{author}{Markus Lumpe}, \bibinfo{author}{Lars Grunske} \&
  \bibinfo{author}{Jean-Guy Schneider} (\bibinfo{year}{2008}):
  \emph{\bibinfo{title}{{State Space Reduction Techniques for Component
  Interfaces}}}.
\newblock In: \bibinfo{editor}{Michel R.~V. Chaudron} \&
  \bibinfo{editor}{Clements Szyperski}, editors: {\sl \bibinfo{booktitle}{CBSE
  2008}}, \bibinfo{series}{LNCS 5282}, \bibinfo{publisher}{Springer},
  \bibinfo{address}{Heidelberg, Germany}, pp. \bibinfo{pages}{130--145}.
\bibitemend

\bibitemstart{lynch:87a}
\bibinfo{author}{Nancy~A. Lunch} \& \bibinfo{author}{Mark R.~Tuttle Tuttle}
  (\bibinfo{year}{1987}): \emph{\bibinfo{title}{{Hierarchical Correctness
  Proofs for Distributed Algorithms}}}.
\newblock In: {\sl \bibinfo{booktitle}{Proceedings of the Sixth Annual ACM
  Symposium on Principles of Distributed Computing}},
  \bibinfo{address}{Vancouver, British Columbia, Canada}, pp.
  \bibinfo{pages}{137--151}.
\bibitemend

\bibitemstart{nash:50a}
\bibinfo{author}{John~Forbes Nash} (\bibinfo{year}{1950}):
  \emph{\bibinfo{title}{{Non-cooperative Games}}}.
\newblock \bibinfo{type}{Ph.D. thesis}, \bibinfo{school}{Department of
  Mathematics, Princeton University}.
\bibitemend

\bibitemstart{newman:03a}
\bibinfo{author}{M.~E.~J. Newman} (\bibinfo{year}{2003}):
  \emph{\bibinfo{title}{{The Structure and Function of Complex Networks}}}.
\newblock {\sl \bibinfo{journal}{SIAM Review}} \bibinfo{volume}{45}, pp.
  \bibinfo{pages}{167--256}.
\bibitemend

\bibitemstart{potanin:05}
\bibinfo{author}{Alex Potanin}, \bibinfo{author}{James Noble},
  \bibinfo{author}{Marcus~R. Frean} \& \bibinfo{author}{Robert Biddle}
  (\bibinfo{year}{2005}): \emph{\bibinfo{title}{{Scale-Free Geometry in OO
  Programs}}}.
\newblock {\sl \bibinfo{journal}{Commun. ACM}}
  \bibinfo{volume}{48}(\bibinfo{number}{5}), pp. \bibinfo{pages}{99--103}.
\bibitemend

\bibitemstart{shannon:01a}
\bibinfo{author}{Claude~E. Shannon} (\bibinfo{year}{2001}):
  \emph{\bibinfo{title}{{A Mathematical Theory of Communication}}}.
\newblock {\sl \bibinfo{journal}{SIGMOBILE Mob. Comput. Commun. Rev.}}
  \bibinfo{volume}{5}(\bibinfo{number}{1}), pp. \bibinfo{pages}{3--55}.
\bibitemend

\bibitemstart{change:06a}
\bibinfo{author}{{The CHANCE Project}} (\bibinfo{year}{2006}):
  \emph{\bibinfo{title}{{Grinstead and Snell's Introduction to Probability}}}.
\newblock \bibinfo{publisher}{The Chance Project, Mathematics Dept., Dartmouth
  College}.
\bibitemend

\bibitemstart{hdr:07a}
\bibinfo{author}{{United Nations Devlopment Programme}} (\bibinfo{year}{2007}).
\newblock \emph{\bibinfo{title}{{Human Development Report 2007/2008}}}.
\bibitemend

\bibitemstart{valverde:03a}
\bibinfo{author}{Sergi Valverde} \& \bibinfo{author}{Ricard~V. Sol{\'e}}
  (\bibinfo{year}{2003}): \emph{\bibinfo{title}{{Hierarchical Small Worlds in
  Software Architecture}}}.
\newblock {\sl \bibinfo{journal}{Arxiv preprint cond-mat/0307278}} .
\bibitemend

\bibitemstart{vasa:09a}
\bibinfo{author}{Rajesh Vasa}, \bibinfo{author}{Markus Lumpe},
  \bibinfo{author}{Philip Branch} \& \bibinfo{author}{Oscar Nierstrasz}
  (\bibinfo{year}{2009}): \emph{\bibinfo{title}{{Comparative Analysis of
  Evolving Software Systems Using the Gini Coefficient}}}.
\newblock In: {\sl \bibinfo{booktitle}{Proceedings of 25th IEEE International
  Conference on Software Maintenance (ICSM '09)}}, \bibinfo{publisher}{IEEE
  Computer Society}, \bibinfo{address}{Edmonton, Alberta}, pp.
  \bibinfo{pages}{179--188}.
\bibitemend

\bibitemstart{vasa:07a}
\bibinfo{author}{Rajesh Vasa}, \bibinfo{author}{Markus Lumpe} \&
  \bibinfo{author}{Jean-Guy Schneider} (\bibinfo{year}{2007}):
  \emph{\bibinfo{title}{{Patterns of Component Evolution}}}.
\newblock In: \bibinfo{editor}{Markus Lumpe} \& \bibinfo{editor}{Wim
  Vanderperren}, editors: {\sl \bibinfo{booktitle}{Proceedings of the 6th
  International Symposium on Software Composition (SC 2007)}},
  \bibinfo{series}{LNCS 4829}, \bibinfo{publisher}{Springer},
  \bibinfo{address}{Heidelberg, Germany}, pp. \bibinfo{pages}{235--251}.
\bibitemend

\bibitemstart{vasa:05a}
\bibinfo{author}{Rajesh Vasa}, \bibinfo{author}{Jean-Guy Schneider},
  \bibinfo{author}{Clinton Woodward} \& \bibinfo{author}{Andrew Cain}
  (\bibinfo{year}{2005}): \emph{\bibinfo{title}{{Detecting Structural Changes
  in Object-Oriented Software Systems}}}.
\newblock In: \bibinfo{editor}{June Verner} \& \bibinfo{editor}{Guilherme~H.
  Travassos}, editors: {\sl \bibinfo{booktitle}{Proceedings of 4th
  International Symposium on Empirical Software Engineering (ISESE '05)}},
  \bibinfo{publisher}{IEEE Computer Society Press}, \bibinfo{address}{Noosa
  Heads, Australia}, pp. \bibinfo{pages}{463--470}.
\bibitemend

\bibliographyend

\end{thebibliography}

\end{document}